\documentclass[11pt]{article}
\usepackage[a4paper,margin=25mm]{geometry}
\usepackage[T1]{fontenc}
\usepackage[utf8]{inputenc}
\usepackage{lmodern}
\usepackage{microtype}
\usepackage{amsmath,amssymb,mathrsfs,bm}
\usepackage{array}
\usepackage{xcolor}
\usepackage[hidelinks]{hyperref}

\hypersetup{pdftitle={Black holes and scalar propagation in three-dimensional Einstein-Gauss-Bonnet gravity},pdfauthor={Cendikiawan Suryaatmadja}}
\allowdisplaybreaks[2]
\title{Black holes and scalar propagation in three-dimensional Einstein--Gauss--Bonnet gravity}
\author{
Cendikiawan Suryaatmadja%
\thanks{\texttt{cendikiawan.suryaatmadja@uwaterloo.ca}}\\
{\small Department of Physics and Astronomy, University of Waterloo}\\
{\small Waterloo, ON N2L 3G1, Canada}
}
\date{}
\begin{document}
\maketitle
\begin{abstract}
We derive an analytic black-hole family in three-dimensional scalar--tensor Einstein--Gauss--Bonnet gravity with positive coupling. A single implicit equation determines the static circular metric and a scalar linear in time. We show that these solutions exhaust regular nonextremal exteriors with the stated AdS boundary conditions and a fixed nonzero coefficient $q$ in $\phi=qt+\psi(r)$. The coupled metric and scalar perturbations have one propagating degree of freedom. On one branch, its kinetic coefficient is positive, its equation is hyperbolic throughout the exterior, and the spatial part of its quadratic bulk energy is positive for perturbations of compact support. Scalar signals can cross the metric horizon outward, so exterior evolution needs information from the interior. For linear perturbations with the original metric and scalar boundary values fixed, nonzero, compactly supported initial displacements of the master field with zero initial velocity can evolve only until their first contact with the AdS boundary. We derive this restriction from the original metric and scalar equations.
\end{abstract}

\section{Introduction}\label{sec:1}

Three-dimensional Einstein--Gauss--Bonnet gravity has a scalar--tensor formulation with derivative interactions~\cite{LuPang2020,Hennigar2020BTZ,MaLu2020}. A known static family has $\phi=\log(r/\ell)$, and a rotating family follows from a boost and an angular identification~\cite{Hennigar2021Rotating}. We allow $\phi=qt+\psi(r)$ while keeping the metric static and circular. The scalar changes the background equations and the propagation of perturbations, even though three-dimensional Einstein gravity has no propagating gravitational wave.

We classify the regular nonextremal exteriors in this scalar ansatz, derive their coupled metric and scalar perturbations, and determine the restrictions imposed by the original AdS boundary values.

Scalars linear in time also occur in other three-dimensional Horndeski theories~\cite{BravoGaeteHassaine2014}. For the present theory, the Birkhoff-type result in Ref.~\cite{Fernandes2025} assumes a radial logarithmic scalar. The three-dimensional no-news result of Ref.~\cite{LuMao2021} assumes asymptotic flatness and a decaying scalar. Neither set of boundary conditions includes the solutions studied here. The perturbations in Ref.~\cite{CuadrosMelgar2022} are scalar and spinor probes on a fixed geometry; our calculation varies both fields in the gravitational action.

We use signature $(-++)$, $G_N=1$, $\alpha>0$, and $\ell>0$. For zero internal curvature, the action derived in Ref.~\cite{LuPang2020} reduces in three dimensions to the model studied in Refs.~\cite{Hennigar2020BTZ,Hennigar2021Rotating}:
\begin{equation}
\begin{aligned}
I=\frac1{16\pi}\int d^3x\sqrt{-g}\bigg[
R+\frac2{\ell^2}+\alpha\Big(
4G^{ab}\nabla_a\phi\nabla_b\phi
-4(\nabla\phi)^2\Box\phi
+2\big[(\nabla\phi)^2\big]^2\Big)\bigg].
\end{aligned}
\label{eq:1}
\end{equation}
We write $X=(\nabla\phi)^2$, with no factor of $-1/2$. The bare cosmological constant is $\Lambda=-1/\ell^2$. On the branch that approaches Einstein gravity as $\alpha\to0$, the effective cosmological constant and AdS length are
\begin{equation}
\begin{gathered}
\Lambda_\alpha=-\frac1{\ell_\alpha^2}
=-\frac1{\ell^2}\frac{2}{\sqrt{1+4\alpha/\ell^2}+1},\\
\Lambda_\alpha-\alpha\Lambda_\alpha^2=\Lambda,
\qquad B_\infty=1-2\alpha\Lambda_\alpha=\sqrt{1+4\alpha/\ell^2}.
\end{gathered}
\label{eq:2}
\end{equation}
Thus $\Lambda_\alpha<0$ and $\Lambda_\alpha\to\Lambda$ as $\alpha\to0$; $B_\infty$ is the asymptotic value of $B$ defined below.

\section{Field equations and static black holes}\label{sec:2}
\subsection{Field equations}

The scalar gradient and Hessian enter the field equations through
\begin{equation}
\begin{gathered}
\phi_a=\nabla_a\phi,\qquad 
C_{ab}=\nabla_a\nabla_b\phi+\phi_a\phi_b-Xg_{ab},\qquad C=C^a{}_a,\\
Q_{ab}=C_{ac}C^c{}_b-C\,C_{ab}
+\frac12g_{ab}\left(C^2-C_{cd}C^{cd}\right),\qquad B=1+2\alpha X.
\end{gathered}
\label{eq:3}
\end{equation}

Varying every metric component and the scalar before choosing coordinates gives
\begin{equation}
E_{ab}=BG_{ab}-(\ell^{-2}+\alpha X^2)g_{ab}+4\alpha Q_{ab}=0,
\label{eq:4}
\end{equation}
\begin{equation}
\begin{gathered}
J^a=8\alpha\left[G^{ab}\phi_b+(X-\Box\phi)\phi^a+\frac12\nabla^aX\right],
\qquad E_\phi=-\nabla_aJ^a,\\
\frac{E_\phi}{8\alpha}
=C^2-C_{ab}C^{ab}+(Xg^{ab}-G^{ab})C_{ab}=0.
\end{gathered}
\label{eq:5}
\end{equation}

The current $J^a$ in Eq.~\eqref{eq:5} follows from constant scalar-shift symmetry. The three-dimensional relation between the Riemann and Ricci tensors gives the metric equation in the form of Eq.~\eqref{eq:4}. Coordinate invariance implies
\begin{equation}
2\nabla^aE_{ab}+E_\phi\nabla_b\phi=0.
\label{eq:7}
\end{equation}

\subsection{The black-hole family and its assumptions}

Fix $q>0$, $\phi_0$, $\alpha$, $\ell$, the boundary time normalization, and a $2\pi$-periodic angular coordinate. For every $m\in\mathbb R$, there is a unique positive real analytic function $X(r)$ on $r>0$ satisfying
\begin{equation}
\mathcal H(r,X)
=r^2(X+\alpha X^2-\ell^{-2})
-\frac{q^2}{X}+2\alpha q^2\log(X\ell_\alpha^2)+m=0.
\label{eq:8}
\end{equation}

The fields
\begin{equation}
\begin{gathered}
F=r^2X-\frac{q^2}{X},\qquad
ds^2=-Fdt^2+\frac{dr^2}{F}+r^2d\theta^2,
\qquad \phi=qt+\psi(r),\qquad
\psi_{r}=\frac{r X}{F}
\end{gathered}
\label{eq:9}
\end{equation}

solve Eqs.~\eqref{eq:4}--\eqref{eq:5}. Each has one nonextremal metric horizon, and the scalar is smooth across its future horizon.

For the classification, assume a connected static circular exterior with $\phi=qt+\psi$ throughout, positive lapse and circumference, and fields of class $C^3$ at every finite exterior point. The metric and scalar are smooth in coordinates crossing a nonextremal future horizon, with finite $X$ there. Near infinity, use the circumference $2\pi r$ to define $r$, and require
\begin{equation}
\begin{gathered}
g_{tt}=-\frac{r^2}{\ell_\alpha^2
}+O(1),\qquad
 g_{rr}=\frac{\ell_\alpha^2}{ r^2}+O(r^{-4}),\\
\phi=qt+\log(r/\ell_\alpha)+O(r^{-2})+\textrm{const.}.
\end{gathered}
\label{eq:asymptotic_assumptions}
\end{equation}
The expansions may be differentiated twice in $r$. The boundary metric, time normalization, coefficient of $\log r$, constants $q\ne0,\phi_0$, and $2\pi$ angular period are fixed. Under these assumptions, Eqs.~\eqref{eq:8}--\eqref{eq:9} give all solutions. The proof also establishes that $r$ is a valid coordinate throughout the exterior and that the radial lapse is constant.

\subsection{Branch selection}

Keep the radial lapse and the off-diagonal metric components independent when varying Eq.~\eqref{eq:1} to obtain $E_{ab}=0$ in Eq.~\eqref{eq:4}. Then choose proper distance $\rho$ in the exterior, with $\rho=0$ at the horizon:
\begin{equation}
\begin{gathered}
ds^2=-\mathcal N(\rho)^2dt^2+d\rho^2+r(\rho)^2d\theta^2,
\\
a_\rho=\frac{\dot{\mathcal N}}{\mathcal N},\quad b_\rho=\frac{\dot{r}}{r},
\quad X=\dot\psi^2-q^2/\mathcal N^2,\quad U_*=X-b_\rho \dot\psi.
\end{gathered}
\label{eq:10}
\end{equation}

Dots denote $d/d\rho$. In the orthonormal frame of Eq.~\eqref{eq:10}, the components $E_{01}$ and $E_{11}$ of Eq.~\eqref{eq:4} are
\begin{equation}
E_{01}=\frac{4\alpha q}{\mathcal N}(\dot\psi-a_\rho)U_*,\qquad
E_{11}=Ba_\rho b_\rho-\ell^{-2}-\alpha X^2+4\alpha \dot\psi(\dot\psi-a_\rho)U_*.
\label{eq:11}
\end{equation}

Since $q\ne0$, $E_{01}=0$ in Eq.~\eqref{eq:11} gives $(\dot\psi-a_\rho)U_*=0$. Substituting this into $E_{11}=0$ yields
\begin{equation}
B a_\rho b_\rho=\ell^{-2}+\alpha X^2>0.
\end{equation}
This relation holds without choosing which factor, $\dot\psi-a_\rho$ or $U_*$, vanishes.

At every regular exterior point, $a_\rho$ and $b_\rho$ are finite,
so the positive right-hand side excludes $B=0$. The asymptotic
conditions Eq.~\eqref{eq:asymptotic_assumptions} give
$X=\ell_\alpha^{-2}+O(r^{-2})$, and hence
$B\to B_\infty=1+2\alpha/\ell_\alpha^2>0$.
Since $B$ is continuous and nowhere zero, it follows that $B>0$
throughout the connected exterior. Hence $a_\rho b_\rho>0$,
and neither factor can vanish.

Near the nonextremal horizon, $\mathcal N\sim\kappa_h\rho$, so $a_\rho\sim1/\rho>0$. The positive product then gives $b_\rho>0$ there. Neither quantity can change sign without passing through zero, so both remain positive throughout the exterior. In particular, $\dot r=b_\rho r>0$, and the circumference is strictly increasing.

The scalar derivative $\dot\psi$ cannot vanish either. If $\dot\psi=0$ at an exterior point, then $U_*=-q^2/\mathcal N^2\ne0$, and $E_{01}=0$ would force $a_\rho=0$, contradicting the result above. Equation~\eqref{eq:asymptotic_assumptions} gives $\partial_r\psi=1/r+O(r^{-3})>0$ near infinity. Since $\dot\psi=(\partial_r\psi)\dot r$ and $\dot r>0$, it follows that $\dot\psi>0$ there; continuity then gives $\dot\psi>0$ throughout the exterior. Finally, smoothness across the future horizon requires $\dot\psi=q/(\kappa_h\rho)+O(\rho)$, where $\kappa_h>0$ is the surface gravity. Since $\dot\psi>0$, this requires $q>0$.

The components $E_{00}$ and $E_{22}$ in the same frame are
\begin{equation}
\begin{aligned}
E_{00}&=-B(\dot{b_\rho}+b_\rho^2)+\ell^{-2}+\alpha X^2
+4\alpha U_*(\ddot\psi+q^2/\mathcal N^2),\\
E_{22}&=B(\dot{a_\rho}+a_\rho^2)-\ell^{-2}-\alpha X^2
-2\alpha(\dot\psi-a_\rho)\dot X.
\end{aligned}
\label{eq:12}
\end{equation}

Set $D=(\ell^{-2}-X-\alpha X^2)/B$. Suppose that $U_*$ is nonzero somewhere, and consider a maximal open radial interval on which $U_*\ne0$. $E_{01}=0$ in Eq.~\eqref{eq:11} then requires $\dot\psi=a_\rho$. The equation $E_{11}=0$ gives $\dot\psi b_\rho=X+D$, while $E_{22}=0$ in Eq.~\eqref{eq:12} gives $\ddot\psi+\dot\psi^2=X+D$. Using the definitions of $U_*$ and $X$ in Eq.~\eqref{eq:10}, we obtain
\begin{align}
U_*&=X-\dot\psi b_\rho=-D,\\
\dot X&=2\dot\psi\left(\ddot\psi+\frac{q^2}{\mathcal N^2}\right)
=2\dot\psi D.
\end{align}

Since $\dot{\mathcal N}/\mathcal N=\dot\psi$ on this interval,
\begin{equation}
\frac{d}{d\rho}\left[\mathcal N^2(X+\alpha X^2-\ell^{-2})\right]
=\mathcal N^2B(\dot X-2\dot\psi D)=0.
\end{equation}
Thus $\mathcal N^2(X+\alpha X^2-\ell^{-2})=-\mathcal N^2BD$ is constant. It is nonzero because $D=-U_*\ne0$ and $\mathcal N,B>0$ on the interval.

At an endpoint lying in the regular exterior, continuity and maximality would require $U_*=0$. The relation $D=-U_*$ would then make the constant vanish, a contradiction. The interval therefore has no endpoint in the regular exterior and must extend to the horizon.

At the horizon, however, $\mathcal N\to0$ and $X$ remains finite, so the same constant tends to zero. This is again a contradiction. There can therefore be no interval with $U_*\ne0$, and $U_*=0$ throughout the exterior.

Equations~\eqref{eq:11}--\eqref{eq:12} now give
\begin{equation}
X=b_\rho \dot\psi>0,\qquad \dot{b_\rho}+b_\rho^2=a_\rho b_\rho,\qquad
\frac{d}{d\rho}\log\frac{\mathcal N}{b_\rho r}=0.
\label{eq:13}
\end{equation}

Since $b_\rho>0$, the circumference increases outward, so $r$ can be used throughout the exterior. The asymptotic forms of $g_{tt}$ and $g_{rr}$ in Eq.~\eqref{eq:asymptotic_assumptions} fix $\mathcal N=b_\rho r$. Writing primes for $d/dr$, $E_{ab}=0$ in Eq.~\eqref{eq:4} and $U_*=0$ give
\begin{equation}
F=r^2X-\frac{q^2}{X},\qquad
BF'=2r(\ell^{-2}+\alpha X^2),\qquad
X'=\frac{2r X^2D}{r^2X^2+q^2}.
\label{eq:14}
\end{equation}

Eq.~\eqref{eq:8} is a first integral. At $X=\ell_\alpha^{-2}$, uniqueness for the equation for $X'$ in Eq.~\eqref{eq:14} prevents a constant solution and a nonconstant solution from joining at a finite radius.

\subsection{Global solution and horizon}

For each fixed $r>0$, the function $\mathcal H(r,X)$ defined in
Eq.~\eqref{eq:8} satisfies
\begin{equation}
\frac{\partial\mathcal H}{\partial X}=B\left(r^2+\frac{q^2}{X^2}\right)>0,
\qquad
\lim_{X\to0^+}\mathcal H=-\infty,
\qquad
\lim_{X\to\infty}\mathcal H=+\infty.
\label{eq:15}
\end{equation}

For every $r>0$, Eq.~\eqref{eq:15} gives exactly one positive root. Since $\frac{\partial\mathcal H}{\partial X}\ne0$, that root is analytic in $r$. A horizon satisfies $X_h=q/r_h$. Equivalently,
\begin{equation}
m(r_h)=\ell^{-2}r_h^2-\alpha q^2
-2\alpha q^2\log\frac{q\ell_\alpha^2}{r_h},
\qquad
\frac{dm}{dr_h}=2\ell^{-2}r_h+\frac{2\alpha q^2}{r_h}>0.
\label{eq:16}
\end{equation}

This function runs from $-\infty$ to $+\infty$. Thus every real $m$ gives one positive horizon radius. Eq.~\eqref{eq:14} gives $F'>0$, so this is the only horizon, it is nonextremal, and the entire outer region has $F>0$.

Choose advanced Eddington--Finkelstein (EF) time with $v-t\to0$ at infinity. The scalar with boundary offset $\phi_0$ is
\begin{equation}
\begin{aligned}
ds^2&=-Fdv^2+2dv\,dr+r^2d\theta^2,\\
\phi&=qv+\log\frac{r}{\ell}+\phi_0
+\int_{r}^{\infty}
\frac{q\,d\rho}{\rho[\rho X(\rho)+q]},\\
\partial_{r}\phi\big|_v&=\frac{X}{r X+q},
\qquad
\partial_{r}\phi\big|_{v,h}=\frac1{2r_h}.
\end{aligned}
\label{eq:17}
\end{equation}

The integral converges, and its denominator remains positive through the future horizon. The metric and scalar are analytic in $(v,r,\theta)$ across the future horizon.

For the fields in Eq.~\eqref{eq:9}, the definitions in Eq.~\eqref{eq:3} and the relations in Eq.~\eqref{eq:14} give
\begin{equation}
C^\theta{}_{\theta}=0,\qquad
C=\frac{2r^2X^2D}{r^2X^2+q^2},\qquad
\frac{Q_{\theta\theta}}{r^2}=\frac{r^2X^2D^2}{r^2X^2+q^2}.
\notag
\end{equation}
In the static coordinates of Eq.~\eqref{eq:9}, the equations $E_{tt}=E_{tr}=E_{rr}=0$ from Eq.~\eqref{eq:4} reduce to Eq.~\eqref{eq:14}. The equation $E_{\theta\theta}=0$ follows by differentiating the relations in Eq.~\eqref{eq:14}. In $E_\phi=0$ in Eq.~\eqref{eq:5}, $C^2-C_{ab}C^{ab}=2r^2X^2D^2/(r^2X^2+q^2)=DC$, which cancels $(Xg^{ab}-G^{ab})C_{ab}=-DC$.

Define
\begin{equation}
d_0=\frac{m-q^2\ell_\alpha^2}{B_\infty}.
\label{eq:18}
\end{equation}

The asymptotic metric and scalar in static time are
\begin{equation}
\begin{gathered}
X=\ell_\alpha^{-2}-\frac{d_0}{r^2}+O(r^{-4}),\qquad
F=\frac{r^2}{\ell_\alpha^2}-d_0-q^2\ell_\alpha^2+O(r^{-2}),\\
\phi=qt+\log\frac{r}{\ell}+\phi_0
-\frac{q^2\ell_\alpha^4}{2r^2}+O(r^{-4}).
\end{gathered}
\label{eq:19}
\end{equation}

Since $\mathcal H(r,\ell_\alpha^{-2})=m-q^2\ell_\alpha^2$ and $\frac{\partial\mathcal H}{\partial X}>0$, the sign of $D$ is fixed throughout the solution by $m-q^2\ell_\alpha^2$. The perturbation analysis below concerns $d_0>0$, equivalently $0<X<\ell_\alpha^{-2}$ and $D>0$. Section~\ref{sec:3} shows that this branch has a positive scalar kinetic coefficient. At $d_0=0$, $X=\ell_\alpha^{-2}$ everywhere and the scalar quadratic action degenerates. For $d_0<0$, the scalar is asymptotically a ghost. At fixed parameters the metric tends to BTZ as $\alpha\to0$; on the $d_0>0$ branch the limiting BTZ horizon has positive radius.

\subsection{Conserved charges}

The potentials $\Theta^a$ and $Q_\xi^{ab}$ are given in Eq.~\eqref{eq:20} in Appendix~\ref{app:A}. At fixed $q,\phi_0,\alpha,\ell$, they give
\begin{equation}
\delta H_{\partial_t}(r)
=\frac{\delta m}{8}+\frac{\alpha q^2}{2}\delta\log X(r),
\qquad \delta M_\infty=\frac{\delta m}{8}.
\label{eq:21}
\end{equation}

Thus the mass changes by $\delta m/8$ when the boundary values are held fixed. The additive reference energy depends on the boundary prescription. Evaluating $J^a$ from Eq.~\eqref{eq:5} gives
\begin{equation}
J^{r}=0,\qquad
r J^t=4\alpha q(\log X)',\qquad
Q_{\rm shift}[r_1,r_2]
=\frac{\alpha q}{2}\log\frac{X(r_2)}{X(r_1)}.
\label{eq:22}
\end{equation}

The scalar changes under time translation, $\mathcal L_{\partial_t}\phi=q$, so $\partial_t$ alone is not a symmetry of both fields. The radial change of Eq.~\eqref{eq:21} is $q\,\delta Q_{\rm shift}[r_1,r_2]$; the mass is defined at infinity.

The full background is invariant under a time translation accompanied by a constant scalar shift, $\delta\phi=\mathcal L_{\partial_t}\phi-q=0$. At fixed $q$, this transformation acts on perturbations as $\partial_t=\partial_\tau$. The mass variation $\delta M_\infty=\delta m/8$ describes changes within the black-hole family; the quadratic energy derived below from the action for the master field describes perturbations about a fixed background. Its conservation also depends on flux through the radial boundaries.

For nonconstant $X$, the shift current is nonzero. The known radial logarithmic family has vanishing shift current, which remains zero under a coordinate boost. A boost of that solution cannot give the nonconstant-$X$ family in Eq.~\eqref{eq:8}. In the class studied here, $q$ and $\phi_0$ are fixed boundary data, and $m$ is the only free parameter of the black hole. The related vector--tensor theory of Ref.~\cite{Alkac2025Proca} reduces to this scalar action under $W_a=-\nabla_a\phi$ with their vector coupling set to $-\alpha$. Independent vector variation would require $J^a=0$; Eq.~\eqref{eq:22} therefore distinguishes the present family from that restriction.

\section{Coupled perturbations}\label{sec:3}

\subsection{Propagation and the sign of the kinetic term}

The metric and scalar perturbations are coupled. Solving the constraints in $\delta E_{ab}=0$, with $\delta E_{ab}$ given in Eq.~\eqref{eq:full-metric-linearization}, leaves one scalar degree of freedom. We use $I^{(2)}$ for the coefficient of the squared perturbation parameter in the action, with $h_{ab}=\delta g_{ab}$ and $\pi=\delta\phi$. Appendix~\ref{app:A} derives the reduction by varying all metric components before choosing a gauge. The scalar principal tensor and the part of the quadratic action containing two derivatives are
\begin{equation}
Z^{ab}=2(Cg^{ab}-C^{ab})-Dg^{ab}+\frac{12\alpha}{B}Q^{ab},
\qquad
I^{(2)}_{\rm pr}=-\frac{4\alpha}{16\pi}\int d^3x\sqrt{-g}
\,Z^{ab}\partial_a\pi\partial_b\pi.
\label{eq:25}
\end{equation}

For the solution in Eq.~\eqref{eq:9}, let $y_*=q/(r X)$. In the advanced chart,
\begin{equation}
\begin{aligned}
Z^{vv}&=-\frac{4Dy_*}{r^2X(1+y_*)^2(1+y_*^2)},&
Z^{vr}&=D\frac{1-y_*}{1+y_*},\\
Z^{rr}&=Dr^2X(1+y_*^2),&
Z^{\theta\theta}&=\frac D{r^2}
\frac{3B_\infty^2/B^2-y_*^2}{1+y_*^2}.
\end{aligned}
\label{eq:26}
\end{equation}

On the $D>0$ branch, the scalar has a positive kinetic coefficient and its equation is hyperbolic at every finite exterior point and across the future metric horizon. The function $T=v-\Delta v(r)$ has a timelike gradient $T_a=\partial_aT$ for both the metric and scalar principal tensor, where
\begin{equation}
(\Delta v)'=\frac1{r^2X(1+y_*)(1+y_*^2)},\qquad
Z^{ab}T_aT_b=-\frac{D(1+2y_*)}{r^2X(1+y_*)^2(1+y_*^2)}<0.
\label{eq:27}
\end{equation}

Substitution also gives $g^{ab}T_aT_b<0$, so constant-$T$ surfaces are spacelike for both systems. In the exterior $0<y_*\le1$, so $Z^{\theta\theta}>0$ in Eq.~\eqref{eq:26}. The outgoing and ingoing radial scalar rays obey, respectively,
\begin{equation}
\frac{dr}{dv}
=\frac{r^2X(1+y_*)(1+y_*^2)}2,
\qquad
\frac{dr}{dv}
=-\frac{r^2X(1+y_*)(1+y_*^2)}{2y_*}.
\label{eq:28}
\end{equation}

At $r_hX_h=q$, their radial coordinate speeds are $\pm2qr_h$. One scalar characteristic therefore crosses the metric horizon outward. Regularity at the metric horizon does not determine the incoming scalar data. An exterior evolution problem must specify this input, either by evolving interior data or by prescribing the incoming scalar data at a stated inner boundary. Appendix~\ref{app:D} gives the energy balance and the region that remains unaffected by this interior input.

At $D=0$, $C_{ab}=0$, and the scalar terms in the quadratic action, including the metric--scalar mixing, vanish. Eq.~\eqref{eq:24} starts at quadratic order in perturbations and remains nontrivial. The loss of the scalar kinetic term therefore signals strong coupling. For $D<0$, the scalar kinetic term has the opposite sign near infinity. On the $D>0$ branch, $D\sim d_0/r^2$ tends to zero at infinity, so the asymptotic evolution problem requires a separate analysis of the boundary conditions.

At any finite exterior radius on this branch, dividing the scalar principal operator $Z^{ab}\nabla_a\nabla_b$, with $Z^{ab}$ given in Eq.~\eqref{eq:25}, by $D>0$ preserves its characteristic cone. The principal part of the quadratic action in Eq.~\eqref{eq:25} retains the factor $D$, so positivity at finite radius does not give a uniform lower bound on the kinetic coefficient towards infinity. In optical coordinates the radial weight in the action for each nonzero angular harmonic is $w=\frac{d_0x}{\ell_\alpha^2}[1+O(x^2)]$: scalar signals reach the boundary in finite optical time although this weight vanishes there. The asymptotic degeneracy requires specifying the allowed boundary behaviour of the linear solutions. By itself it neither establishes a nonlinear singularity nor determines a strong-coupling scale.

\subsection{Circular sector}\label{sec:4}

The following combinations simplify the perturbation equations:
\begin{equation}
S=r^2X^2+q^2,\qquad A=rX+q,\qquad p=\frac XA,
\qquad X'=\frac{2rX^2D}{S},\qquad D'=-\left(1+\frac{2\alpha D}{B}\right)X'.
\label{eq:29}
\end{equation}
Here $p=\partial_r\phi|_v$ is the scalar derivative at fixed advanced time. In the circular sector even under $\theta\to-\theta$, choose the radial coordinate so that $h_{\theta\theta}=0$, and use the advanced time of Eq.~\eqref{eq:17}:
\begin{equation}
\begin{gathered}
h_{vv}=-(w_c+2Fn),\qquad h_{vr}=n,\qquad
h_{rr}=h_{\theta\theta}=0,\qquad \delta\phi=\pi,\\
U_c=Bw_c=B\{\delta[(\nabla r)^2]-F'\delta r\}.
\end{gathered}
\label{eq:30}
\end{equation}
The last expression defines the gauge-invariant field before fixing the areal radius. The angular-momentum perturbation vanishes in this parity sector, and a change of mass gives $U_c=-\delta m$. Eq.~\eqref{eq:33} in Appendix~\ref{app:A} recovers $n$ and $\pi$ from $U_c$; the argument following it verifies the remaining equations $\delta E_{\theta\theta}=0$ and $\delta E_\phi=0$, obtained by linearizing Eqs.~\eqref{eq:4}--\eqref{eq:5}.

Choose coordinates in which radial scalar signals have unit coordinate speed:
\begin{equation}
\tau=v-\int^r z(\rho)d\rho,\qquad
z=\frac{X(rX-q)}{AS},\qquad
x(r)=\int_r^\infty\frac{X(\rho)}{\rho^2X(\rho)^2+q^2}\,d\rho.
\label{eq:35}
\end{equation}
Here $x=0$ is the AdS boundary, $x_h=x(r_h)$, and $x$ increases inward. The additive time constant is chosen so that $\tau-t\to0$ at infinity. Constant-$\tau$ slices remain spacelike for the scalar across the horizon; the common spacelike slices for the metric and scalar are given in Eq.~\eqref{eq:27}. The exact circular bulk action and equation are
\begin{equation}
\begin{gathered}
I_c^{(2)}=\frac1{16q}\int d\tau\,dx\,W_c(U_{c,\tau}^2-U_{c,x}^2),
\qquad W_c=\frac1{8\alpha Dqr},\\
U_{c,\tau\tau}-W_c^{-1}(W_cU_{c,x})_x=0.
\end{gathered}
\label{eq:36}
\end{equation}
On a radial interval $0<x<x_i$, where $x_i=x(r_i)$ is the chosen inner endpoint, the positive bulk energy on $D>0$ obeys
\begin{equation}
E_c=\frac1{16q}\int_0^{x_i} W_c(U_{c,\tau}^2+U_{c,x}^2)dx,
\qquad
\frac{dE_c}{d\tau}=\left[\frac{W_c}{8q}U_{c,\tau}U_{c,x}\right]_{0}^{x_i}.
\label{eq:37}
\end{equation}
This is the energy obtained from the bulk action for $U_c$. Additional boundary contributions depend on the boundary conditions and boundary action.

\subsection{Angular sector}\label{sec:5}

For each angular harmonic $e^{ik\theta}$, with integer $k\ne0$, one gauge-invariant field determines both the scalar and metric perturbations. The scalar amplitude $P$ and metric amplitude $V$ are defined in Eq.~\eqref{eq:65} of Appendix~\ref{app:A}; the gauge-invariant field is
\begin{equation}
\mathcal U=P+\frac{ikp}{K}V,\qquad K=k^2+q^2/X+r^2D>0.
\notag
\end{equation}
Eq.~\eqref{eq:71} recovers $h_{ab}$ and $\pi$, including time-independent perturbations, without a singular denominator at the horizon. Write $U(\tau,x)=\mathcal U(v,r)$. Its equation is
\begin{equation}
U_{\tau\tau}-w^{-1}(wU_x)_x+\mathcal V_kU=0,
\qquad w=rD>0.
\label{eq:41}
\end{equation}
For a real angular harmonic normalized to unit $L^2$ norm, matching the symplectic current to that of the original metric--scalar action, as shown in Eq.~\eqref{eq:43}, fixes the overall coefficient of the quadratic bulk action:
\begin{equation}
I_k^{(2)}=\frac{\alpha}{4\pi}\int d\tau\,dx\,w
\left(U_\tau^2-U_x^2-\mathcal V_kU^2\right).
\label{eq:42}
\end{equation}
Completing the square in the spatial part gives
\begin{equation}
\begin{gathered}
\widehat\nu=\frac{4\alpha Dr(Xk^2+q^2)}{BK},\\
\widehat P_k=k^2\left[
\frac F{r^2}+2X+\frac{4\alpha DX}{B}
+\frac{8\alpha D(Xk^2+q^2)}{BK}\right]>0,\\
\mathcal V_k=\widehat\nu^2+\widehat\nu_x+(\log w)_x\widehat\nu+\widehat P_k,\\
\int w(|U_x|^2+\mathcal V_k|U|^2)dx
=\int w(|U_x-\widehat\nu U|^2+\widehat P_k|U|^2)dx
+[w\widehat\nu|U|^2]_{\rm endpoints}.
\end{gathered}
\label{eq:44}
\end{equation}

Since $\widehat P_k>0$ throughout the exterior and at the horizon, Eq.~\eqref{eq:44} proves positivity of the spatial part of the bulk energy when the perturbation vanishes outside a finite radial interval. For general boundary values, the displayed boundary term must also be included.

To separate the angular propagation speed from lower-derivative terms, write
\begin{equation}
\begin{gathered}
\mathcal V_k=c_\theta^2k^2+\mathcal V_{k,\mathrm{rem}},\qquad
c_\theta^2=\frac F{r^2}+2X+\frac{12\alpha DX}{B}>0,\\
\mathcal V_{k,\mathrm{rem}}=\widehat\nu^2+\widehat\nu_x+(\log w)_x\widehat\nu
-\frac{8\alpha D^2Xr^2}{B}\frac{k^2}{K}.
\end{gathered}
\label{eq:45}
\end{equation}

Using $B^2+4\alpha BD=B_\infty^2$, the speed $c_\theta$ agrees with the characteristic tensor in Eq.~\eqref{eq:26}. On $0\le x\le x_h$, every fixed number of $x$ derivatives of $\mathcal V_{k,\mathrm{rem}}$ is bounded independently of $|k|\ge1$. This follows from $k^2/K\le1$ and Eq.~\eqref{eq:49}. Consequently $\mathcal V_{k,\mathrm{rem}}$ does not require control of additional angular derivatives in the energy estimate in Eq.~\eqref{eq:58}.

\section{AdS sources and evolution}\label{sec:6}

We fix the conformal metric $-ds^2+d\theta^2$, where $s=t/\ell_\alpha$, the logarithmic coefficient of $\phi$, and its finite boundary profile $qt+\phi_0$. We exclude terms proportional to $\log r$ at order $r^0$ in the boundary metric components and impose the expansions and derivative bounds in Appendix~\ref{app:C}. The boundary action and leading volume counterterm are given in Eq.~\eqref{eq:46}. We include no additional finite boundary term. With $\Box_{(0)}=-\partial_s^2+\partial_\theta^2$, temporarily allowing the finite scalar profile to vary gives the renormalized momentum
\begin{equation}
\lim_{r\to\infty}\sqrt{-\gamma}\Pi_\phi
=-\frac{8\alpha}{\ell_\alpha}\,\Box_{(0)}f,
\qquad f=qt+\phi_0\ \Longrightarrow\ \lim_{r\to\infty}\sqrt{-\gamma}\Pi_\phi=0.
\label{eq:48}
\end{equation}
The symplectic flux of the original action through the AdS boundary therefore vanishes when these sources are fixed; see Eqs.~\eqref{eq:74}--\eqref{eq:78}.

\subsection{The scalar source and the boundary value of the master field}\label{sec:7}

For $k\ne0$, define $K_\infty=\lim_{r\to\infty}K=k^2+d_0-q^2/\Lambda_\alpha>0$. The weight in optical coordinates is $w=-\Lambda_\alpha d_0x[1+O(x^2)]$. Both the regular and logarithmic radial solutions have finite $L^2(wdx)$ norm, but only the regular solution has finite radial gradient energy before renormalization. Its boundary value $b_k(\tau)=U_k(\tau,0)$ is related to the scalar source variation $p_{0,k}$ and the finite metric response $a_k$ by
\begin{equation}
\begin{gathered}
b_k=p_{0,k}+\frac{-\Lambda_\alpha k^2+q\partial_\tau}{k^2K_\infty}a_k,
\qquad (\partial_\tau^2-\Lambda_\alpha k^2)a_k=0,\\
p_{0,k}=0\ \Longrightarrow\
(\partial_\tau^2-\Lambda_\alpha k^2)b_k=0,
\qquad
a_k=\frac{K_\infty(\Lambda_\alpha k^2+q\partial_\tau)}{\Lambda_\alpha(q^2-\Lambda_\alpha k^2)}b_k.
\end{gathered}
\label{eq:51}
\end{equation}
For $p_{0,k}=0$, the formula for $a_k$ in Eq.~\eqref{eq:51} inverts the relation between $a_k$ and $b_k$ on solutions of $(\partial_\tau^2-\Lambda_\alpha k^2)b_k=0$. Equations~\eqref{eq:50} and \eqref{eq:79} specify the metric response and the required boundary regularity; Appendix~\ref{app:C} derives their relation to $U_k$. Thus fixing the original scalar source imposes an evolution equation on the boundary value of the master field. The boundary value of the master field need not vanish.

\begin{figure}[t]
\centering

\small
\begin{tabular}{@{}c@{\quad$\longrightarrow$\quad}c@{\quad$\longrightarrow$\quad}c@{}}
\shortstack{Original action\\and reconstructed fields}
&\shortstack{Symplectic current\\Eq.~\eqref{eq:43}}
&\shortstack{Quadratic action\\coefficient $\alpha/(4\pi)$}
\\[1.5em]
\shortstack{Metric constraints\\in FG coordinates}
&\shortstack{Reconstruction of $b_k$\\from $a_k$ and $p_{0,k}$}
&\shortstack{Fixed $p_{0,k}=0$\\Oscillator equation for $b_k$}
\end{tabular}
\caption{Matching the symplectic currents fixes the coefficient $\alpha/(4\pi)$ in the quadratic action. The metric constraints in Eq.~\eqref{eq:50} and $p_{0,k}=0$ imply the oscillator equation for $b_k$ in Eq.~\eqref{eq:51}. Regularity and vanishing energy flux at the AdS boundary do not by themselves impose this equation.}
\label{fig:two-reductions}

\end{figure}

For the circular master field $U_c$, the expansion is
\begin{equation}
\begin{gathered}
W_c=\frac{\ell_\alpha^2}{8\alpha qd_0x}[1+O(x^2)],\\
U_c=u_0(\tau)+x^2\left[u_2(\tau)+\frac12u_{0,\tau\tau}(\tau)\log(x/x_0)\right]+\cdots,\\
\delta\phi_{\log}=-\frac{\ell_\alpha^2}{8\alpha qd_0}u_{0,\tau},\qquad
\partial_\tau(\delta\phi_{\rm finite})=\frac{\ell_\alpha^2}{8\alpha qd_0}\left(2u_2+\frac12u_{0,\tau\tau}\right).
\end{gathered}
\label{eq:52}
\end{equation}
Here $x_0>0$ is a fixed reference length. Setting $\delta\phi_{\log}=0$ in Eq.~\eqref{eq:52} makes $u_0$ constant; this is the mass variation. At fixed mass, $u_0=0$. Setting $\delta\phi_{\rm finite}=0$ then also requires $u_2=0$ and fixes the initial scalar shift.

\subsection{First contact with the boundary}\label{sec:8}

Consider linear perturbations of a $D>0$ background, with a nonzero real smooth initial displacement $U_0$ of the master field, compactly supported in the open exterior, with zero angular mean and \emph{zero initial velocity}, $U_\tau|_{\tau=0}=0$. Equation~\eqref{eq:71} recovers metric and scalar initial perturbations satisfying the constraints in $\delta E_{ab}=0$, with $\delta E_{ab}$ given in Eq.~\eqref{eq:full-metric-linearization}. Let $T_{\max}$ be the upper limit on durations of this linear evolution with the sources fixed as in Section~\ref{sec:6} and the expansions and derivative bounds specified with Eqs.~\eqref{eq:50} and \eqref{eq:79}, allowing a choice of incoming scalar data through the metric horizon. Then
\begin{equation}
T_{\max}=d,
\qquad
d=\int_{r_{\max}}^\infty
\frac{X(\rho)}{\rho^2X(\rho)^2+q^2}\,d\rho,
\qquad 0<d<x_h.
\label{eq:57}
\end{equation}
Here $r_{\max}$ is the outermost radius in the initial support and $d$ is its optical distance to AdS. Solutions exist on every interval $0\le\tau<T<d$, and none exists on such an interval with $T>d$. The equality includes smooth packets that vanish to every order at the edge of their support.

Appendix~\ref{app:D} constructs a smooth comparison solution with an auxiliary Dirichlet boundary just inside the metric horizon. Finite propagation preserves the original sources for $\tau<d$. Evolution beyond $d$ would violate Eq.~\eqref{eq:51}, as proved by the boundary estimate in Eq.~\eqref{eq:55}. The comparison solution can remain smooth after contact: its finite scalar source then fails to remain fixed. For the circular sector of Section~\ref{sec:4}, the same conclusion holds at fixed mass for a nonzero smooth displacement compactly supported in the open exterior, with zero initial velocity, using Eq.~\eqref{eq:52}.

For the angular modes, regularity gives $U_k=b_k(\tau)+O(x^2)$ and $U_{k,x}=O(x)$, so the energy flux of the master field through the AdS boundary vanishes for any smooth $b_k$. Fixing the original scalar source imposes the additional oscillator equation in Eq.~\eqref{eq:51}. Compact initial data give $b_k(0)=\partial_\tau b_k(0)=0$, and the oscillator equation then requires $b_k=0$ throughout the evolution. The uniqueness estimate in Eq.~\eqref{eq:55} excludes continuation beyond the contact time. The prescribed sources therefore prevent continuation within the stated regularity class: the smooth comparison solution continues, but its scalar source changes. The theorem concerns the stated linear initial data, including zero initial velocity; it does not establish a nonlinear instability. Positive bulk energy and vanishing AdS flux alone do not guarantee existence with the original sources.

\label{sec:9}
The combination specifying the incoming scalar data required by Eq.~\eqref{eq:28} has principal part
\begin{equation}
U_\tau+U_x\quad\text{at a specified inner boundary }x=x_i.
\label{eq:61}
\end{equation}
Smooth initial perturbations just inside the metric horizon can send an outward pulse into an otherwise unperturbed exterior. Until that pulse reaches AdS, the same exterior initial fields and fixed sources therefore admit different smooth evolutions.
\section{Discussion}\label{sec:10}

At fixed $q$ and $\phi_0$, the black-hole family has one parameter, $m$, under the assumptions of Section~\ref{sec:2}. The $D>0$ branch has a positive scalar kinetic coefficient and a positive spatial contribution to the bulk energy for perturbations with compact support in the exterior. Stability also depends on boundary conditions and data entering from the interior. Allowing the finite scalar profile to vary, with the term in Eq.~\eqref{eq:117}, defines a different boundary problem.

The analysis covers the exterior and a short extension across the future horizon. Deeper inside, $Z^{\theta\theta}$ in Eq.~\eqref{eq:26} vanishes when $(q/(rX))^2=3B_\infty^2/B^2$. Evolution through that surface, nonlinear existence, and long-term stability of the complete spacetime remain open.

\appendix

\section{Recovering the metric and scalar perturbations}\label{app:A}

We recover $h_{ab}$ and $\pi$ from the master fields, verify the linearized equations $\delta E_{ab}=0$ and $\delta E_\phi=0$ obtained from Eqs.~\eqref{eq:4}--\eqref{eq:5}, and fix the bulk action coefficients.

\subsection{Metric variation before gauge fixing}

We first vary all six metric components and the scalar independently. With $h_{ab}=\delta g_{ab}$, $\pi=\delta\phi$, and $h^{ab}=g^{ac}g^{bd}h_{cd}$,
\begin{equation}
\delta I=\frac1{16\pi}\int_M d^3x\sqrt{-g}
\left(-E^{ab}h_{ab}+E_\phi\pi\right)
+\frac1{16\pi}\int_{\partial M}d\Sigma_a\,\Theta^a.
\end{equation}
The potential $\Theta^a$ is given in Eq.~\eqref{eq:20}. No component of $h_{ab}$ is set to zero in this variation. The quantities in Eq.~\eqref{eq:4} vary as
\begin{align}
\delta X&=2\phi^a\nabla_a\pi-h^{ab}\phi_a\phi_b,\\
\delta\Gamma^c{}_{ab}&=\frac12g^{cd}
(\nabla_a h_{bd}+\nabla_b h_{ad}-\nabla_d h_{ab}),\\
\delta C_{ab}&=\nabla_a\nabla_b\pi-\delta\Gamma^c{}_{ab}\phi_c
+\phi_a\nabla_b\pi+\phi_b\nabla_a\pi-g_{ab}\delta X-Xh_{ab},\\
\delta C&=g^{ab}\delta C_{ab}-h^{ab}C_{ab},\\
\delta Q_{ab}&=C_a{}^c\delta C_{cb}+C_b{}^c\delta C_{ac}
-h^{cd}C_{ac}C_{bd}-C_{ab}\delta C-C\delta C_{ab}\notag\\
&\quad+\frac12h_{ab}(C^2-C_{cd}C^{cd})
+g_{ab}\left(C\delta C-C^{cd}\delta C_{cd}
+h^{cd}C_c{}^eC_{de}\right),\\
\delta E_{ab}&=B\delta G_{ab}+2\alpha\delta X\,G_{ab}
-2\alpha X\delta X\,g_{ab}
-(\ell^{-2}+\alpha X^2)h_{ab}+4\alpha\delta Q_{ab}.
\label{eq:full-metric-linearization}
\end{align}
Here $\delta G_{ab}=\delta R_{ab}-\tfrac12h_{ab}R-\tfrac12g_{ab}\delta R$, where
\begin{equation}
\delta R_{ab}=\nabla_c\delta\Gamma^c{}_{ab}
-\nabla_b\delta\Gamma^c{}_{ac},\qquad
\delta R=g^{ab}\delta R_{ab}-h^{ab}R_{ab}.
\end{equation}
On the background solution, but before imposing any perturbation equation, Eq.~\eqref{eq:7} becomes
\begin{equation}
2\nabla^a\delta E_{ab}+\phi_b\delta E_\phi=0.
\label{eq:linear-noether}
\end{equation}
We impose the coordinate conditions only after deriving $\delta E_{ab}=0$ and $\delta E_\phi=0$. The equations for the metric components fixed by this choice are therefore retained and checked below.

Keeping the terms with two derivatives of the perturbations in Eq.~\eqref{eq:full-metric-linearization} gives
\begin{equation}
\delta E_{ab}=B\,\delta G_{ab}(h)
+8\alpha\,\delta G_{ab}(C\pi)+\text{lower derivative terms}.
\label{eq:23}
\end{equation}
Expressing the metric equation in terms of
$h_{ab}+8\alpha C_{ab}\pi/B$ removes the mixing of second derivatives. Eliminating $G_{ab}$ from $E_\phi=0$ in Eq.~\eqref{eq:5} using Eq.~\eqref{eq:4} gives
\begin{equation}
\begin{aligned}
0={}&B\left(C^2-C_{ab}C^{ab}\right)
+(X+\alpha X^2-\ell^{-2})C\\
&+4\alpha\left[C^a{}_bC^b{}_cC^c{}_a
-\frac32 C\,C_{ab}C^{ab}+\frac12C^3\right].
\end{aligned}
\label{eq:24}
\end{equation}
The cubic bracket is $3\det(C^a{}_b)$, and its derivative with respect to $C_{ab}$ at fixed metric is $3Q^{ab}$. Dividing Eq.~\eqref{eq:24} by $B$ and differentiating with respect to the scalar Hessian therefore gives $Z^{ab}$ in Eq.~\eqref{eq:25}. This elimination uses the metric equations; treating $\pi$ as a probe on a fixed metric would retain a different principal operator.

\subsection{Circular equations}

For the variables in Eq.~\eqref{eq:30}, define
\begin{equation}
\begin{gathered}
\mathsf a=\frac{rX-q}{8\alpha DqrA},\qquad
\mathsf h=\frac{S}{8\alpha DXqr},\qquad
\mathsf j=\frac{X^2}{2\alpha DA^2S}.
\end{gathered}
\label{eq:31}
\end{equation}

The equations $\delta E_{rr}=\delta E_{vr}=\delta E_{vv}=0$, with $\delta E_{ab}$ given in Eq.~\eqref{eq:full-metric-linearization}, give
\begin{equation}
\begin{aligned}
n_r&=\frac{X^2}{BA^2S}U_{c,v},\\
\pi_r&=\frac{qX^2}{BA^2S}U_c+\mathsf a U_{c,r}-\mathsf j U_{c,v},\\
\pi_v-qn&=-\mathsf h U_{c,r}-\mathsf a U_{c,v}.
\end{aligned}
\label{eq:33}
\end{equation}

The condition $\partial_v\pi_r=\partial_r\pi_v$, with $n_r$ eliminated using the first line of Eq.~\eqref{eq:33}, gives
\begin{equation}
\partial_v(-\mathsf j U_{c,v}+\mathsf a U_{c,r})
+\partial_r(\mathsf a U_{c,v}+\mathsf h U_{c,r})=0.
\label{eq:32}
\end{equation}

The coefficients remain finite at $F=0$. With $\delta E_{rr}=\delta E_{vr}=\delta E_{vv}=0$, the $v$ and $r$ components of Eq.~\eqref{eq:linear-noether} give $q\delta E_\phi=0$ and $\delta E_{\theta\theta}=0$. The boundary time and scalar shift fix the remaining integration functions. Eq.~\eqref{eq:35} converts Eq.~\eqref{eq:32} to Eq.~\eqref{eq:36}.

\subsection{Angular equations}

Take $e^{ik\theta}$, $k\ne0$. The time-derivative operators below act on the perturbations to their right. Their coefficients depend only on $r$ and commute with $\partial_v$. A prime denotes radial differentiation at fixed $v$; when applied to an operator, it differentiates its radial coefficients. Define
\begin{equation}
\begin{gathered}
K=k^2+\frac{q^2}{X}+r^2D>0,\qquad
\mathfrak b=-\frac{ikBA}{4\alpha Dr^3S}\ne0,\qquad
\mathfrak w=\frac{rDS}{X},\\
\mu=\frac{Xk^2}{rS}+\frac{4\alpha DXr(Xk^2+q^2)}{BSK},\qquad
\chi=\frac{X^2k^2(F-k^2)}{r^2S^2}.
\end{gathered}
\label{eq:38}
\end{equation}

The radial weights in the $r$ and $x$ coordinates are $\mathfrak w$ and $w=rD$, respectively. The linearized equations $\delta E_{ab}=0$ and $\delta E_\phi=0$, obtained from Eqs.~\eqref{eq:4}--\eqref{eq:5}, reduce to the following system, which is first order in $r$:
\begin{equation}
\mathcal U=P+\frac{ikp}{K}V,\qquad \Pi=\mathfrak bV,
\qquad
\begin{aligned}
\mathcal U_r&=-\mu\mathcal U-z\mathcal U_v+\Pi,\\
\Pi_r&=\chi\mathcal U+\frac{X^2}{S^2}\mathcal U_{vv}
+[\mu-(\log\mathfrak w)']\Pi-z\Pi_v.
\end{aligned}
\label{eq:39}
\end{equation}

Eliminating $\Pi$ and using Eq.~\eqref{eq:35} gives Eq.~\eqref{eq:41}, with
\begin{equation}
\mathcal V_k=-\frac{S^2}{X^2}
\left[\mu'+(\log\mathfrak w)'\mu-\mu^2-\chi\right].
\label{eq:40}
\end{equation}
To recover the metric and scalar, first choose areal advanced gauge:
\begin{equation}
\begin{gathered}
\boldsymbol y=(a,b,d,L,\pi_{\rm EF})^T,\qquad
h_{vv}=-a,\quad h_{vr}=b,\quad h_{v\theta}=d,\quad
\\
h_{rr}=h_{r\theta}=h_{\theta\theta}=0,
\quad \pi_{\rm EF}=\delta\phi,\quad L=rB(d'-2d/r).
\end{gathered}
\label{eq:62}
\end{equation}

The residual coordinate transformations have generator
\begin{equation}
\xi^a=\left(T,\,k^2T-ikr\zeta,\,\frac{ikT}{r}+\zeta\right)e^{ik\theta},
\qquad \partial_rT=\partial_r\zeta=0,
\end{equation}
in the coordinate order $(v,r,\theta)$. Their Lie derivatives act on $\boldsymbol y$ through

\begin{equation}
G_T=\begin{pmatrix}
k^2F'+2F\partial_v-2k^2\partial_v\\ \partial_v\\ ik(r\partial_v-F+k^2)\\
-ikB(r\partial_v+2K)\\q+pk^2
\end{pmatrix},\qquad
G_{\zeta}=\begin{pmatrix}
-ikr(F'-2\partial_v)\\-ik\\r(r\partial_v+k^2)\\-k^2rB\\-ikrp
\end{pmatrix}.
\label{eq:63}
\end{equation}

The determinant of the $(b,L)$ coefficient matrix is $2k^2BK\ne0$. We therefore define combinations satisfying $\widehat b=\widehat L=0$ by
\begin{equation}
T_*=\frac{iL/(kB)-rb}{2K},\qquad
\zeta_*=\frac{\partial_v T_*-b}{ik},\qquad
\widehat{\boldsymbol y}=\boldsymbol y-G_TT_*-G_{\zeta}\zeta_*.
\label{eq:64}
\end{equation}

Equation~\eqref{eq:64} removes $b$ and $L$ at each radius. The resulting $T_*$ and $\zeta_*$ generally depend on $r$, so this subtraction is not itself a residual coordinate transformation preserving areal advanced gauge. Their radial derivatives must be retained. Subtracting the full Lie derivative then gives the radial metric components in Eq.~\eqref{eq:71}.

The metric constraint containing no radial derivatives is $[K+r(\partial_v+q-rD)]\widehat a-2ik(\partial_v+q-rD)\widehat d=0$. It gives
\begin{equation}
\begin{gathered}
P=\widehat\pi_{\rm EF},\qquad V=\widehat d-\frac r{2ik}\widehat a,\qquad
\widehat a=a_VV,\quad\widehat d=d_VV,\\
a_V=\frac{2ik(\partial_v+q-rD)}{K},\qquad d_V=1+\frac{r(\partial_v+q-rD)}{K}.
\end{gathered}
\label{eq:65}
\end{equation}

The equations $\delta E_{rr}=\delta E_{vr}=\delta E_{r\theta}=0$, together with $L=rB(d'-2d/r)$ from Eq.~\eqref{eq:62}, determine the radial derivatives of $P,V,T_*,\zeta_*$. Their coefficients can be written in terms of
\begin{equation}
\begin{gathered}
h_*=\frac{8\alpha DXqr}{BA^2},
\\
C_{vr}=\frac{DrX}{A},\quad C_{rr}=\frac{2DX^2qr}{A^2S},
\quad C=\frac{2r^2X^2D}{S},\\
E_P=-\frac{k^2}{r^2}+\frac{q-rX}{rA}\partial_v,\qquad
E_V=\frac{Xq}{rA^2}a_V-\frac{ikX}{r^2A}d_V.
\end{gathered}
\label{eq:66}
\end{equation}

In $\delta E_{r\theta}=0$, differentiate $L=rB(d'-2d/r)$ to express $d''$ in terms of $L'$ and lower radial derivatives. Retain the defining relation for $L$ as the fourth equation, so the system remains equivalent to the original equations.

Use $r\delta E_{rr}/B$, $2r\delta E_{vr}/B$, $-2\delta E_{r\theta}/(ikB)$, and $d'-2d/r-L/(rB)$. After these substitutions, their coefficients multiplying $P$ and $V$ without radial derivatives are the rows
\begin{equation}
\begin{aligned}
J_1={}&-\frac{4\alpha rC_{rr}}B(E_P,E_V),\\
J_2={}&\left(\frac{8\alpha rDp}B\partial_v,
 a_V'-\frac{ik}{r}d_V'+\frac{4\alpha rDp^2a_V}B\right)
 +\frac{8\alpha r(C-C_{vr})}B(E_P,E_V),\\
J_3={}&\left(-\frac{8\alpha}B[C_{rr}(\partial_v+q)-C_{vr}(p-1/r)],
-\frac{4\alpha C_{rr}pa_V}B\right),\\
J_4={}&(0,d_V'-2d_V/r).
\end{aligned}
\label{eq:67}
\end{equation}

Write $W=P'+(q+pk^2)T_*'-ikrp\zeta_*'$. The four equations are
\begin{equation}
\begin{gathered}
\mathsf M
\begin{pmatrix}W\\ V'\\ T_*'\\ \zeta_*'\end{pmatrix}
=-\begin{pmatrix}J_1\\J_2\\J_3\\J_4\end{pmatrix}
\begin{pmatrix}P\\V\end{pmatrix},\\
\mathsf M=\begin{pmatrix}
h_*&0&\partial_v&-ik\\
-Fh_*&a_V-ikd_V/r&k^2(F'-\partial_v-F/r+k^2/r)&ik(-rF'+r\partial_v+2F-k^2)\\
\frac{A}{q}h_*&0&2\partial_v+2K/r&-2ik\\
0&d_V&ik(r\partial_v-F+k^2)&r(r\partial_v+k^2)
\end{pmatrix}.
\end{gathered}
\label{eq:68}
\end{equation}
Its determinant is
\begin{equation}
\det\mathsf M=\frac{16i\alpha DkS\widetilde K}{BA}\ne0,
\qquad
\widetilde K=k^2+\frac{q^2}{X}+\frac{2Dq^2r^2}{S}>0.
\label{eq:69}
\end{equation}
The quantities $B,A,S,K,\widetilde K,D$ are positive throughout the stated domain, including the metric horizon. Eq.~\eqref{eq:68} therefore determines the radial derivatives also for time-independent perturbations. After solving this system, the operators mapping $(P,V)$ to $T_*'$ and $\zeta_*'$ have orders at most one and two in $\partial_v$, respectively. For a solution of Eq.~\eqref{eq:41}, set
\begin{equation}
\Pi=\mathcal U_r+\mu\mathcal U+z\mathcal U_v,
\qquad V=\Pi/\mathfrak b,\qquad P=\mathcal U-\frac{ikp}{K}V.
\label{eq:70}
\end{equation}

Solve Eq.~\eqref{eq:68} for $T_*',\zeta_*'$. Integrating the resulting equations for $T_*$ and $\zeta_*$ in Eqs.~\eqref{eq:63}--\eqref{eq:65}, then subtracting the full Lie derivative with generator $(T_*,k^2T_*-ikr\zeta_*,ikT_*/r+\zeta_*)$, cancels the integrals and gives
\begin{equation}
\begin{aligned}
\pi&=P,& h_{\theta\theta}&=0,\\
h_{vv}&=-\frac{2ik}{K}(\partial_v+q-rD)V,
&h_{v\theta}&=\left[1+\frac rK(\partial_v+q-rD)\right]V,\\
h_{rr}&=-2T_*',
&h_{vr}&=(F-k^2)T_*'+ikr\zeta_*',\\
h_{r\theta}&=-ikrT_*'-r^2\zeta_*'.
\end{aligned}
\label{eq:71}
\end{equation}

Eq.~\eqref{eq:68} imposes $\delta E_{rr}=\delta E_{vr}=\delta E_{r\theta}=0$. The constraint used in Eq.~\eqref{eq:65} gives $\delta E_{v\theta}=ikrp\,\delta E_{vv}/q$. The components of $\nabla_a\delta E^a{}_b$ are
\begin{equation}
\begin{aligned}
\nabla_a\delta E^a{}_{\theta}
 &=\partial_r\delta E_{v\theta}+\frac{\delta E_{v\theta}}r
   +\frac{ik}{r^2}\delta E_{\theta\theta},\\
\nabla_a\delta E^a{}_v
 &=\partial_r\delta E_{vv}+\frac{\delta E_{vv}}r
   +\frac{ik}{r^2}\delta E_{v\theta},\\
\nabla_a\delta E^a{}_r&=-\frac{\delta E_{\theta\theta}}{r^3}.
\end{aligned}
\label{eq:72}
\end{equation}

Substitute $\delta E_{v\theta}=ikrp\,\delta E_{vv}/q$ into the $\theta$ component and the $v-(q/p)r$ combination of Eq.~\eqref{eq:linear-noether}. They give, respectively,
\begin{align}
0&=\partial_r\delta E_{vv}
+\left(\frac2r+\frac{p'}p\right)\delta E_{vv}
+\frac{q}{r^3p}\delta E_{\theta\theta},\\
0&=\partial_r\delta E_{vv}
+\left(\frac1r-\frac{k^2p}{qr}\right)\delta E_{vv}
+\frac{q}{r^3p}\delta E_{\theta\theta}.
\end{align}
Subtracting gives
\begin{equation}
0=\left(\frac1r+\frac{p'}p+\frac{k^2p}{qr}\right)\delta E_{vv}
=\frac{p\widetilde K}{qr}\delta E_{vv}.
\end{equation}
It follows that $\delta E_{vv}=\delta E_{v\theta}=\delta E_{\theta\theta}=0$. Eq.~\eqref{eq:linear-noether} then gives $\delta E_\phi=0$. Together with Eq.~\eqref{eq:68}, these verify $\delta E_{ab}=0$ and $\delta E_\phi=0$, including at zero temporal frequency.

The expressions \eqref{eq:70}--\eqref{eq:71} use at most three derivatives of the master field. For initial values $U_0,U_1$, put $\mathcal A_k=-w^{-1}\partial_x(w\partial_x)+\mathcal V_k$. Eq.~\eqref{eq:41} gives
\begin{equation}
\left.\partial_\tau^{2j}U\right|_0=(-\mathcal A_k)^jU_0,
\qquad
\left.\partial_\tau^{2j+1}U\right|_0=(-\mathcal A_k)^jU_1.
\notag
\end{equation}
These expressions vanish wherever both initial functions vanish on an open radial interval. The same is therefore true of $h_{ab},\pi$ and their initial time derivatives. On the initial slice $\tau=0$, and any closed interval $J$ at finite radius with $D>0$, the reconstructed fields obey
\begin{equation}
\begin{split}
&\|h_{ab}\|_{H^N(J)}+\|\pi\|_{H^N(J)}
+\|\partial_\tau h_{ab}\|_{H^N(J)}+\|\partial_\tau\pi\|_{H^N(J)}\\
&\qquad\le C_{N,J}(1+|k|)^8
\left(\|U_0\|_{H^{N+4}(J)}+\|U_1\|_{H^{N+3}(J)}\right).
\end{split}
\label{eq:reconstruction-bound}
\end{equation}
Here $H^N$ measures square-integrable radial derivatives through order $N$; metric components are summed. Eq.~\eqref{eq:69} bounds the denominators. Eq.~\eqref{eq:71} contributes at most four powers of $|k|$, and replacing time derivatives by $\mathcal A_k$ adds at most four more, by Eq.~\eqref{eq:45}. Radial differentiation preserves these bounds, so smooth Fourier sums reconstruct smooth fields. At infinity the coefficients grow by fixed powers of $r$; fields vanishing faster than every power of $r^{-1}$ retain this property.

\subsection{Normalization from the original action}

With $h_{ab}=\delta g_{ab}$ and $h=g^{ab}h_{ab}$, the curvature momentum $P^{abcd}$, symplectic potential $\Theta^a$, and Noether potential $Q_\xi^{ab}$ are
\begin{equation}
\begin{aligned}
P^{abcd}&=\frac{1-2\alpha X}{2}(g^{ac}g^{bd}-g^{ad}g^{bc})\\
&\quad+\alpha(g^{ac}\phi^b\phi^d-g^{ad}\phi^b\phi^c-g^{bc}\phi^a\phi^d+g^{bd}\phi^a\phi^c),\\
\Theta^a&=2P^{abcd}\nabla_dh_{bc}-2\nabla_dP^{abcd}h_{bc}
+J^a\delta\phi-4\alpha X\nabla^a\delta\phi\\
&\quad+4\alpha X\phi^ch^a{}_c-2\alpha X\phi^ah,\\
Q_\xi^{ab}&=-2P^{abcd}\nabla_c\xi_d+4\xi_d\nabla_cP^{abcd}
-4\alpha X(\phi^ag^{bc}-\phi^bg^{ac})\xi_c.
\end{aligned}
\label{eq:20}
\end{equation}

Eq.~\eqref{eq:20} omits the common factor $1/(16\pi)$. The symplectic current density is $\boldsymbol\omega^a=[\delta_1(\sqrt{-g}\Theta^a[\delta_2])-\delta_2(\sqrt{-g}\Theta^a[\delta_1])]/(16\pi)$. On a background solution, this current has vanishing divergence when both perturbations satisfy $\delta E_{ab}=0$ and $\delta E_\phi=0$. It depends on two perturbations and is distinct from the scalar-shift current $J^a$. The potential $Q_\xi^{ab}$, together with $\Theta^a$, gives the mass variation \eqref{eq:21}.

For the circular sector, put $z_c=\pi_v-qn$ and use $U_c=Bw_c$. Expanding Eq.~\eqref{eq:1} to quadratic order, integrating over $\theta$, and integrating by parts gives
\begin{equation}
\begin{gathered}
I_H^{(2)}=\frac18\int dv\,dr\,[U_c n_r+L_0],\\
\begin{aligned}
L_0={}&-4\alpha r\left(Z^{vv}z_c^2+2Z^{vr}z_c\pi_r+Z^{rr}\pi_r^2\right)\\
&-4\alpha r C_{rr}w_c\left[
\left(\frac1r-2p\right)z_c-\frac S{rX}\pi_r
+\frac{qX}{2rA^2}w_c\right],
\end{aligned}
\end{gathered}
\label{eq:89}
\end{equation}

Use $Z^{ab}$ from Eq.~\eqref{eq:26} and $C_{rr}$ from Eq.~\eqref{eq:66}. Substituting Eq.~\eqref{eq:33} gives $L_{0,z_c}=-U_{c,r}/q$, $L_{0,\pi_r}=U_{c,v}/q$. Thus the radial symplectic current obtained from Eq.~\eqref{eq:89} is
\begin{equation}
8j_H^r=U_{c,1}n_2-U_{c,2}n_1
 +\frac{U_{c,1,v}\pi_2-U_{c,2,v}\pi_1}{q}.
\notag
\end{equation}
For perturbations proportional to $e^{\sigma v}$ and $e^{-\sigma v}$, Eq.~\eqref{eq:33} reduces it to
\begin{equation}
8qj_H^r=\mathsf h(U_{c,1}U_{c,2,r}-U_{c,1,r}U_{c,2})
-2\sigma\mathsf a U_{c,1}U_{c,2}.
\label{eq:91}
\end{equation}
This is also the radial symplectic current of
$I_c^{(2)}=(16q)^{-1}\int dv\,dr\,
(\mathsf jU_{c,v}^2-2\mathsf aU_{c,v}U_{c,r}-\mathsf hU_{c,r}^2)$.
The integrations by parts in Eq.~\eqref{eq:89} change the original radial current only by a total $v$ derivative, which vanishes for this pairing. Since $\mathsf a/\mathsf h=z$ and $\mathsf hX/S=W_c$, Eq.~\eqref{eq:35} gives the coefficient $1/(16q)$ in Eq.~\eqref{eq:36}.

For angular perturbations, let $j^r_{\rm orig}=\int d\theta\,\boldsymbol\omega^r$ and take
$e^{\sigma v+ik\theta}/\sqrt{2\pi}$ and
$e^{-\sigma v-ik\theta}/\sqrt{2\pi}$. 
The angular integral of their product is one, so the current is
\begin{equation}
\begin{aligned}
16\pi j^r_{\rm orig}=r\bigg(&
\delta_1\Theta^r[\delta_2]-\delta_2\Theta^r[\delta_1]+\frac12\left(h_1\Theta^r[\delta_2]
-h_2\Theta^r[\delta_1]\right)\bigg),
\end{aligned}
\end{equation}
Here $h_i=g^{ab}h_{i,ab}$ is the trace of the $i$th metric perturbation, and the angular factors have already been integrated. The last two terms come from varying $\sqrt{-g}$, whose background value is $r$. Use $\delta g^{ab}=-h^{ab}$, $\delta\phi^a=\nabla^a\pi-h^{ab}\phi_b$, and the variations of $X$, $\Gamma^c{}_{ab}$, and $C_{ab}$ displayed before Eq.~\eqref{eq:full-metric-linearization} when varying Eq.~\eqref{eq:20}. Substituting the fields reconstructed in Eqs.~\eqref{eq:70}--\eqref{eq:71} and eliminating radial derivatives with Eq.~\eqref{eq:68} gives

\begin{equation}
\begin{aligned}
16\pi j^r_{\rm orig}
&=\frac{2ikBA}{r^2X}(P_1V_2+V_1P_2)\\
&=8\alpha\mathfrak w
\left[\mathcal U_1\mathcal U'_2-\mathcal U'_1\mathcal U_2
-2\sigma z\mathcal U_1\mathcal U_2\right].
\end{aligned}
\label{eq:43}
\end{equation}

For these two modes, $\mathfrak b_2=-\mathfrak b_1$, whereas $\mu$ and $z$ are unchanged. Writing $\mathfrak b=\mathfrak b_1$, Eq.~\eqref{eq:70} gives
\begin{align}
P_1V_2+V_1P_2
&=-\frac{1}{\mathfrak b}
(\mathcal U_1\Pi_2-\Pi_1\mathcal U_2),\\
\mathcal U_1\Pi_2-\Pi_1\mathcal U_2
&=\mathcal U_1\mathcal U'_2-\mathcal U'_1\mathcal U_2
-2\sigma z\mathcal U_1\mathcal U_2.
\end{align}
Since $-2ikBA/(r^2X\mathfrak b)=8\alpha\mathfrak w$, these identities give the second line of Eq.~\eqref{eq:43}.

To check the action coefficient, use $x'=-X/S$ and $\tau_r=-z$ to express Eq.~\eqref{eq:42} in $(v,r)$ coordinates. Denoting its density in these coordinates by $\mathcal L_k^{(2)}$, the radial momentum is
\begin{equation}
\frac{\partial\mathcal L_k^{(2)}}{\partial\mathcal U_r}
=-\frac{\alpha}{2\pi}\mathfrak w
(\mathcal U_r+z\mathcal U_v).
\end{equation}
The corresponding radial symplectic current is
\begin{equation}
j_k^r=\frac{\alpha}{2\pi}\mathfrak w
\left[\mathcal U_1\mathcal U'_2-\mathcal U'_1\mathcal U_2
-2\sigma z\mathcal U_1\mathcal U_2\right]
=j_{\rm orig}^r.
\end{equation}
Multiplying the master action by a constant would leave its equation unchanged but rescale this current. Agreement with the original current therefore fixes the coefficient $\alpha/(4\pi)$, including its sign. Total $v$ and $\theta$ derivatives vanish for these two modes because their exponential factors cancel. Boundary contributions are determined separately from the boundary variation; the AdS calculation is given in Eqs.~\eqref{eq:74}--\eqref{eq:78}.

\newpage
\section{Original boundary variation and scalar source}\label{app:C}

\subsection{Boundary action and flux}

Let $n$ be the outward spacelike unit normal, $\gamma_{ij}$ the induced metric, and $D_i$ its covariant derivative. Write $\mathsf K_{ij}=\tfrac12\mathcal L_n\gamma_{ij}$, $\mathsf K=\gamma^{ij}\mathsf K_{ij}$, $\phi_n=n^a\nabla_a\phi$, $\phi_i=D_i\phi$, and $\phi^i=\gamma^{ij}\phi_j$. The boundary action that cancels variations of the normal derivatives of the metric and scalar is
\begin{equation}
\begin{aligned}
I_\partial&=\frac1{16\pi}\int_{\partial M}d^2x\sqrt{-\gamma}
\left(\mathcal B_D+c_0\right),\\
\mathcal B_D&=2\mathsf K+4\alpha\left[
\mathsf K_{ij}D^i\phi D^j\phi-\mathsf K \phi_i\phi^i
+\phi_n \phi_i\phi^i+\frac{\phi_n^3}{3}\right],\\
c_0&=-\frac2{\ell_\alpha}\left(1+\frac{2\alpha}{3\ell_\alpha^2}\right).
\end{aligned}
\label{eq:46}
\end{equation}
This is the Horndeski boundary construction of Padilla and Sivanesan~\cite{PadillaSivanesan2012}, with the leading volume counterterm $c_0$. We vary the metric and scalar before imposing Gaussian normal gauge $h_{nn}=h_{ni}=0$. Write $h=\gamma^{ij}h_{ij}$ and $M^{ij}=\mathsf K^{ij}-\mathsf K\gamma^{ij}+\phi_n\gamma^{ij}$. The scalar boundary momentum is
\begin{equation}
\Pi_\phi=n_aJ^a-8\alpha D_i(M^{ij}D_j\phi).
\label{eq:47}
\end{equation}
Let $\boldsymbol\Theta_{\rm orig}^{n}$ denote the normal component of the symplectic potential density obtained from Eq.~\eqref{eq:20}, including the factor $1/(16\pi)$. Adding the variation of the boundary action gives
\begin{equation}
\begin{gathered}
\boldsymbol\Theta_{\rm orig}^{n}
+\delta\!\left[\frac{\sqrt{-\gamma}}{16\pi}(\mathcal B_D+c_0)\right]
=\mathcal P^{ij}\delta\gamma_{ij}
+\frac{\sqrt{-\gamma}}{16\pi}\Pi_\phi\,\pi+\partial_i\mathcal C_D^i,\\
16\pi\mathcal C_D^i=\sqrt{-\gamma}
\left[2\alpha \phi_n(\phi^jh_j{}^i-\phi^ih)+8\alpha M^{ij}\phi_j\pi\right].
\end{gathered}
\label{eq:74}
\end{equation}

Here $\mathcal P^{ij}$ is the metric boundary momentum. It can be obtained without using the field equations. After adding the boundary term $\mathcal B_D$ and integrating by parts in Gaussian normal coordinates, the bulk density is $\sqrt{-g}\mathcal L_D/(16\pi)$, up to tangential divergences, where
\begin{align}
\mathcal L_D={}&(1-2\alpha\phi_n^2)R[\gamma]
+B(\mathsf K^2-\mathsf K_{ij}\mathsf K^{ij})
+\frac2{\ell^2}
-8\alpha\phi_n M^{ij}D_iD_j\phi\notag\\
&-8\alpha\phi_n\mathsf K_{ij}\phi^i\phi^j
-\frac{8\alpha}{3}\mathsf K\phi_n^3
-4\alpha \phi_i\phi^i D^2\phi+2\alpha X^2,
\qquad X=\phi_n^2+\phi_i\phi^i.
\label{eq:radial-lagrangian}
\end{align}
Here $D^2\phi=\gamma^{ij}D_iD_j\phi$. The normal derivatives of $\mathsf K_{ij}$ and $\phi_n$ have been integrated by parts; their boundary terms are cancelled by $\mathcal B_D$. The two-dimensional identity
\begin{equation}
\mathsf K_i{}^k\mathsf K_{kj}
=\mathsf K\mathsf K_{ij}
-\frac12(\mathsf K^2-\mathsf K_{kl}\mathsf K^{kl})\gamma_{ij}
\end{equation}
combines the scalar-gradient contributions multiplying $\mathsf K^2-\mathsf K_{ij}\mathsf K^{ij}$ into $B=1+2\alpha X$.

Since $\mathsf K_{ij}=\tfrac12\partial_n\gamma_{ij}$, differentiating Eq.~\eqref{eq:radial-lagrangian} gives
\begin{align}
\frac{16\pi}{\sqrt{-\gamma}}\mathcal P^{ij}
={}&B(\mathsf K\gamma^{ij}-\mathsf K^{ij})
-4\alpha\phi_n
\left(D^iD^j\phi-\gamma^{ij}D^2\phi+\phi^i\phi^j\right)\notag\\
&-\frac{4\alpha}{3}\phi_n^3\gamma^{ij}
+\frac{c_0}{2}\gamma^{ij}.
\label{eq:metric-boundary-momentum}
\end{align}
Differentiating $\mathcal L_D$ in Eq.~\eqref{eq:radial-lagrangian} with respect to $\phi_n$ gives
\begin{align}
\Pi_\phi={}&4\alpha\phi_n
\left(\mathsf K^2-\mathsf K_{ij}\mathsf K^{ij}-R[\gamma]\right)
+8\alpha\phi_n(X-\mathsf K\phi_n-2D^2\phi)\notag\\
&-8\alpha\mathsf K_{ij}\phi^i\phi^j
-8\alpha(\mathsf K^{ij}-\mathsf K\gamma^{ij})D_iD_j\phi.
\end{align}
Using
\begin{equation}
G_{nn}=\frac12\left(
\mathsf K^2-\mathsf K_{ij}\mathsf K^{ij}-R[\gamma]\right),
\qquad G_{ni}=D_j\mathsf K^j{}_i-D_i\mathsf K,
\end{equation}
and expanding the tangential divergence in Eq.~\eqref{eq:47} gives the same scalar momentum.

The tangential divergence in Eq.~\eqref{eq:74} follows from the potential in Eq.~\eqref{eq:20}. Integrating the terms containing derivatives of $h_{ij}$ by parts along the boundary gives $2\alpha\phi_n(\phi^jh_j{}^i-\phi^ih)$ in $\mathcal C_D^i$. The scalar term $8\alpha M^{ij}\phi_jD_i\pi$ gives the other term in $\mathcal C_D^i$. The coefficient of $\partial_n\pi$ in the bulk potential is $-4\alpha X$; varying $4\alpha(\phi_n\phi_i\phi^i+\phi_n^3/3)$ contributes $+4\alpha X\partial_n\pi$ and cancels it. These are off-shell identities. Gaussian normal gauge is used to evaluate the boundary momenta after varying the metric and scalar; it does not replace the normal metric equations $E_{nn}=0$ and $E_{ni}=0$ obtained from Eq.~\eqref{eq:4}.

Set $z_{\rm FG}=\ell^{-2}e^{-2\eta}$, with $g_{\eta\eta}=\ell_\alpha^2$ and $r\sim\ell e^\eta$. Thus $\eta$ is dimensionless and $z_{\rm FG}\sim r^{-2}$. For the fixed conformal metric and logarithmic scalar coefficient,
\begin{equation}
\begin{gathered}
\phi^i,h_j{}^i,h=O(z_{\rm FG}),\qquad M^{ij}=O(z_{\rm FG}^2),\qquad
\sqrt{-\gamma}=O(z_{\rm FG}^{-1}),\\
\mathcal C_D^i=O(z_{\rm FG}),\qquad
\delta_1\mathcal C_D^i[\delta_2]-\delta_2\mathcal C_D^i[\delta_1]=O(z_{\rm FG}).
\end{gathered}
\label{eq:75}
\end{equation}
The leading terms $\mathsf K^i{}_j=\ell_\alpha^{-1}\delta^i{}_j+O(z_{\rm FG})$ and $\phi_n=\ell_\alpha^{-1}+O(z_{\rm FG})$ cancel in $M^{ij}$. 
In Eq.~\eqref{eq:metric-boundary-momentum}, the coefficient of the leading isotropic term is
\begin{equation}
\frac{B_\infty}{\ell_\alpha}
-\frac{4\alpha}{3\ell_\alpha^3}
+\frac{c_0}{2}=0.
\end{equation}
After this cancellation, the right-hand side of Eq.~\eqref{eq:metric-boundary-momentum} is $O(z_{\rm FG}^2)$. Hence $\mathcal P^{ij}=O(z_{\rm FG})$ and $\mathcal P^{ij}\delta\gamma_{ij}=O(z_{\rm FG})$ for $\delta\gamma_{ij}=O(1)$. This cancellation holds before imposing a boundary equation on the finite scalar variation.

To evaluate the finite limit of $\sqrt{-\gamma}\Pi_\phi$, write $\phi=\eta+f+b_{(2)}z_{\rm FG}+O(z_{\rm FG}^2)$ and $\gamma_{ij}=z_{\rm FG}^{-1}g_{(0)ij}+g_{(2)ij}+O(z_{\rm FG})$, where $g_{(0)}=-ds^2+d\theta^2$. Indices on these coefficients are raised with $g_{(0)}$. The required terms are
\begin{equation}
\begin{gathered}
X=\ell_\alpha^{-2}
+z_{\rm FG}\left[(D_{(0)}f)^2-\frac{4b_{(2)}}{\ell_\alpha^2}\right]
+O(z_{\rm FG}^2),\\
\Box\phi=\frac2{\ell_\alpha^2}
+z_{\rm FG}\left(\Box_{(0)}f-\frac{g_{(2)}{}^i{}_i}{\ell_\alpha^2}\right)
+O(z_{\rm FG}^2),\\
G_{nn}=\ell_\alpha^{-2}
-\frac{g_{(2)}{}^i{}_i}{\ell_\alpha^2}z_{\rm FG}
+O(z_{\rm FG}^2),\qquad
n^a\nabla_a\phi=\ell_\alpha^{-1}(1-2b_{(2)}z_{\rm FG})+O(z_{\rm FG}^2),\\
\tfrac12n^a\nabla_aX=-\ell_\alpha^{-1}z_{\rm FG}
\left[(D_{(0)}f)^2-\frac{4b_{(2)}}{\ell_\alpha^2}\right]
+O(z_{\rm FG}^2),\qquad
G_{ni}D^i\phi=O(z_{\rm FG}^2).
\end{gathered}
\notag
\end{equation}
Substituting these expansions into $J^a$ in Eq.~\eqref{eq:5} and contracting with $n_a$ gives $n_aJ^a=-8\alpha\ell_\alpha^{-1} z_{\rm FG}\Box_{(0)}f+O(z_{\rm FG}^2)$: the terms containing $g_{(2)}{}^i{}_i$, $b_{(2)}$, and $(D_{(0)}f)^2$ cancel. Multiplication by $\sqrt{-\gamma}$ makes the divergence term in Eq.~\eqref{eq:47} $O(z_{\rm FG})$, so it does not contribute to the limit in Eq.~\eqref{eq:48}.

The resulting scalar contribution to the variation of the quadratic action at the AdS boundary is
\begin{equation}
\delta(I+I_\partial)^{(2)}\big|_{\rm AdS}
=-\frac{\alpha}{2\pi\ell_\alpha}\int ds\,d\theta\,
(\Box_{(0)}p_0)\,\delta p_0.
\label{eq:76}
\end{equation}
Antisymmetrizing Eq.~\eqref{eq:74} in two perturbations cancels the second variation of the boundary action. The tangential divergence tends to zero by Eq.~\eqref{eq:75}. The symplectic current of the original action therefore has the boundary limit
\begin{equation}
\lim_{z_{\rm FG}\to0}\boldsymbol\omega_{\rm orig}^{n}(1,2)
=-\frac{\alpha}{2\pi\ell_\alpha}
\left[(\Box_{(0)}p_{01})p_{02}-(\Box_{(0)}p_{02})p_{01}\right].
\label{eq:77}
\end{equation}
For the fixed scalar source of Section~\ref{sec:6}, $p_{01}=p_{02}=0$, giving
\begin{equation}
\lim_{z_{\rm FG}\to0}\boldsymbol\omega_{\rm orig}^{n}=0.
\label{eq:78}
\end{equation}
The limits hold uniformly on compact time intervals. In particular, the symplectic current evaluated on a perturbation and its time derivative gives zero flux through the AdS boundary. Any contribution to the energy from an inner boundary must be treated separately.

Allowing the finite scalar source to vary and adding the following finite intrinsic boundary term changes Eq.~\eqref{eq:76} to
\begin{equation}
\begin{gathered}
I_{\rm fin}=\frac{c_\partial}{16\pi}\lim_{r_o\to\infty}
\int_{r=r_o}d^2x\sqrt{-\gamma}\,\gamma^{ij}D_i\phi D_j\phi,\\
\delta(I+I_\partial+I_{\rm fin})^{(2)}\big|_{\rm AdS}
=-\frac{4\alpha/\ell_\alpha+c_\partial}{8\pi}
\int ds\,d\theta\,(\Box_{(0)}p_0)\delta p_0.
\end{gathered}
\label{eq:117}
\end{equation}
For $c_\partial=-4\alpha/\ell_\alpha$, the quadratic scalar boundary variation vanishes without imposing an equation on $p_0$. This changes the boundary conditions. The result in Eq.~\eqref{eq:57} concerns evolution with fixed sources and makes no statement about evolution with this choice.

\subsection{Converting the boundary expansion to the master field}

In Fefferman--Graham coordinates, $g_{\eta i}=0$ and the tangential coordinates are $(s,\theta)$. For each nonzero angular harmonic, the assumed expansion and the leading terms of the linearized field equations take the form
\begin{equation}
\begin{gathered}
\delta g_{ij}=\delta g_{(2)ij}(t)+O_4(z_{\rm FG}),\qquad
\delta\phi=p_0(t)+p_1(t)z_{\rm FG}+O_4(z_{\rm FG}^2),\\
\delta g_{(2)}=\begin{pmatrix}a&j_{(0)}\\j_{(0)}&a\end{pmatrix},\qquad
-\ell_\alpha a_t+ikj_{(0)}=0,\qquad
-\ell_\alpha\partial_tj_{(0)}+ika=0,\\
p_1=\frac{\partial_t^2-3\Lambda_\alpha k^2-4q\partial_t}{4\Lambda_\alpha^2}p_0+\frac a2.
\end{gathered}
\label{eq:50}
\end{equation}
The coefficients are $C^4$. The notation $O_4$ means that the stated order holds after up to two $\eta$ derivatives and four time derivatives, uniformly on compact time intervals. 
To obtain the constraints in Eq.~\eqref{eq:50}, initially leave the three components of $\delta g_{(2)ij}$ independent. The components $\delta E_{\eta\eta}$ and $\delta E_{\eta i}$ in Eq.~\eqref{eq:full-metric-linearization} have the leading expansions
\begin{align}
\delta E_{\eta\eta}
&=-B_\infty z_{\rm FG}\,\delta g_{(2)}{}^i{}_i+O(z_{\rm FG}^2),\\
\delta E_{\eta i}
&=-B_\infty z_{\rm FG}
\left(D_{(0)}^j\delta g_{(2)ji}-D_{(0)i}\delta g_{(2)}{}^j{}_j\right)
+O(z_{\rm FG}^2).
\end{align}
Thus $\delta g_{(2)ij}$ is traceless and conserved, giving the matrix and the two first-order equations in Eq.~\eqref{eq:50}. Eliminating $j_{(0)}$ yields
\begin{equation}
(\partial_t^2-\Lambda_\alpha k^2)a=0.
\label{eq:FG-response-equation}
\end{equation}
After imposing $\delta g_{(2)}{}^i{}_i=0$ and $D_{(0)}^j\delta g_{(2)ji}=0$, linearizing the scalar expression $E_\phi/(8\alpha)$ in Eq.~\eqref{eq:5} gives
\begin{equation}
\begin{aligned}
\delta\!\left(\frac{E_\phi}{8\alpha}\right)
={}&-d_0\Lambda_\alpha z_{\rm FG}^2
\left[4p_1-2a-
\frac{\partial_t^2-3\Lambda_\alpha k^2-4q\partial_t}
{\Lambda_\alpha^2}p_0\right]\\
&+O(z_{\rm FG}^3).
\end{aligned}
\end{equation}
Since $d_0>0$, setting $\delta E_\phi=0$ gives the last line of Eq.~\eqref{eq:50}. These equations hold before setting $p_0=0$. The scalar response is therefore constrained by the metric equations as well as by the scalar equation.

To impose areal advanced gauge, choose the coordinate adjustment with integration constants set to zero at the boundary. Its leading terms are
\begin{equation}
\xi^r=-\frac a{2r},\qquad
\xi^v=-\frac a{6\Lambda_\alpha r^3},\qquad
\xi^\theta=-\frac{\ell_\alpha j_{(0)}}{3r^3}.
\notag
\end{equation}
Transforming the metric and scalar perturbations gives the boundary limits
\begin{equation}
\pi_{\rm EF}\to p_0,\qquad a_{\rm EF}\to2\Lambda_\alpha a,\qquad
d_{\rm EF}\to j_{(0)}/\ell_\alpha,\qquad L\to-2B_\infty j_{(0)}/\ell_\alpha.
\notag
\end{equation}
The equation $\delta E_{rr}=0$, with $\delta E_{rr}$ given by Eq.~\eqref{eq:full-metric-linearization}, gives the following equation for $b_{\rm EF}$:
\begin{equation}
b_{\rm EF}'=\frac{4\alpha r C_{rr}}B
\left[-\frac{(\partial_t^2-\Lambda_\alpha k^2)p_0}{2\Lambda_\alpha r^2}+O(r^{-4})\right].
\notag
\end{equation}
Since $C_{rr}=O(r^{-5})$, integrating with $b_{\rm EF}\to0$ as $r\to\infty$ gives $b_{\rm EF}=O(r^{-5})$. Equations~\eqref{eq:64} and \eqref{eq:70} yield
\begin{equation}
\mathcal U=\pi_{\rm EF}-\frac{rp}{2K}(a_{\rm EF}-2Fb_{\rm EF})
+\frac{ikp}{K}d_{\rm EF}-qT_*,
\qquad (T_*)_\infty=-\frac{a_t}{k^2K_\infty}.
\notag
\end{equation}

The constraint $-\ell_\alpha a_t+ikj_{(0)}=0$ in Eq.~\eqref{eq:50} gives $j_{(0)}=\ell_\alpha a_t/(ik)$, so
\begin{equation}
(T_*)_\infty=-\frac{i j_{(0)}}{k\ell_\alpha K_\infty}
=-\frac{a_t}{k^2K_\infty}.
\end{equation}
Also, $rp\to1$, $Fb_{\rm EF}\to0$, and $p\,d_{\rm EF}\to0$. Hence
\begin{equation}
b_k=p_{0,k}-\frac{\Lambda_\alpha}{K_\infty}a_k
+\frac{q}{k^2K_\infty}\partial_t a_k.
\end{equation}
At the boundary $t=\tau$. Applying $\partial_\tau^2-\Lambda_\alpha k^2$ and using Eq.~\eqref{eq:FG-response-equation} gives
\begin{equation}
(\partial_\tau^2-\Lambda_\alpha k^2)(b_k-p_{0,k})=0.
\end{equation}
Setting $p_{0,k}=0$ gives the restriction in Eq.~\eqref{eq:51}. For $a_k$ satisfying Eq.~\eqref{eq:FG-response-equation},
\begin{equation}
(\Lambda_\alpha k^2+q\partial_\tau)
(-\Lambda_\alpha k^2+q\partial_\tau)a_k
=\Lambda_\alpha k^2(q^2-\Lambda_\alpha k^2)a_k,
\end{equation}
which also gives the inverse relation in Eq.~\eqref{eq:51}. The restriction on $b_k$ comes from the original metric constraints and scalar source.
 A residual boundary time transformation preserving all sources would obey both $T_{tt}-\Lambda_\alpha k^2T=0$ and $T_t=qT$; it therefore vanishes for $k\ne0$ and cannot change the source relation.

For the regularity argument, the background gives
\begin{equation}
\begin{gathered}
x=\frac{\ell_\alpha^2}{r}+O(r^{-3}),\qquad
w=\frac{d_0x}{\ell_\alpha^2}[1+O(x^2)],\qquad
\widehat\nu=\nu_1x+O(x^3),\\
\mathcal V_k=\frac{3k^2}{\ell_\alpha^2}+2\nu_1+O(x^2),\\
\nu_1=\frac{4\alpha d_0}{\ell_\alpha^2B_\infty K_\infty}
\left(q^2+\frac{k^2}{\ell_\alpha^2}\right).
\end{gathered}
\label{eq:49}
\end{equation}
Since $X$ is analytic in $r^{-2}$ and $1/r$ is odd and analytic in $x$, the functions $w/x$, $\widehat\nu/x$, and $\mathcal V_k$ are analytic in $x^2$ near AdS. For regular solutions preserving the original sources, these expansions and Eq.~\eqref{eq:51} give
\begin{equation}
U_k=b_k(\tau)+O(x^2),\qquad U_{k,x}=O(x),\qquad
\left(\partial_\tau^2+\frac{k^2}{\ell_\alpha^2}\right)b_k=0.
\label{eq:79}
\end{equation}
For any two regular master solutions, the first two conditions give zero symplectic flux through the AdS boundary when calculated from the reduced action in Eq.~\eqref{eq:42}. They do not require $b_k$ to be fixed. The last condition is an additional consequence of fixing the original scalar source in Eq.~\eqref{eq:51}.

\section{Boundary contact and data from the interior}\label{app:D}

\subsection{Uniqueness from vanishing boundary values}

Consider
\begin{equation}
H_{xx}-H_{\tau\tau}+\frac n xH_x+Q(x)H=0,
\qquad n>0,
\label{eq:53}
\end{equation}
where $Q$ is bounded. Assume continuous boundary values of $H,H_\tau,H_x$, with $H_x(\tau,0)=0$, and enough interior regularity for integration by parts. If $H(\tau,0)=0$ for $|\tau-\tau_c|<T$, then $H=0$ for $x+|\tau-\tau_c|<T$.

To prove this, set $s_\pm=\tau_c\pm(T-x)$ and, for $\mu_0>0$, define
\begin{equation}
E(x)=\frac12\int_{s_-}^{s_+}
\left(|H_x|^2+|H_\tau|^2+\mu_0^2|H|^2\right)d\tau.
\label{eq:54}
\end{equation}
Equation~\eqref{eq:53} and integration by parts give
\begin{equation}
\begin{aligned}
E'(x)={}&-\frac12[|H_x-H_\tau|^2+\mu_0^2|H|^2]_{s_+}
-\frac12[|H_x+H_\tau|^2+\mu_0^2|H|^2]_{s_-}\\
&-\frac n x\int_{s_-}^{s_+}|H_x|^2d\tau
+\operatorname{Re}\int_{s_-}^{s_+}(\mu_0^2-Q)\overline HH_xd\tau
\le CE(x).
\end{aligned}
\label{eq:55}
\end{equation}
The singular term is nonpositive. The boundary assumptions give $E(\epsilon)\to0$, so integration of $E'\le CE$ from $\epsilon$ and then $\epsilon\to0^+$ proves the result on every smaller open triangle.

For Eq.~\eqref{eq:41}, set $H=\sqrt{\frac{\ell_\alpha^2w}{d_0x}}\,U$. Equation~\eqref{eq:49} gives Eq.~\eqref{eq:53} with $n=1$ and bounded
\begin{equation}
Q_k=-\mathcal V_k-\frac12\partial_x^2\log(w/x)
-\frac1{2x}\partial_x\log(w/x)
-\frac14[\partial_x\log(w/x)]^2.
\label{eq:56}
\end{equation}
For Eq.~\eqref{eq:36} at fixed mass, $H=\sqrt{\frac{d_0x}{\ell_\alpha^2Dr}}\,\frac{U_c}{x^2}$ gives $n=3$ and bounded $Q$. The coefficients have even expansions in $x$; these are regular radial equations in auxiliary spaces of dimensions two and four. Smooth radial solutions have continuous $H,H_\tau,H_x$ at $x=0$ and satisfy $H_x(\tau,0)=0$, as required for the estimate in Eq.~\eqref{eq:55}. Finite bulk energy alone would not suffice.

\subsection{Proof of the contact time}

For the lower bound in Eq.~\eqref{eq:57}, continue the background a short distance inside the metric horizon, with $D>0$ and positive radial and angular spatial coefficients, and impose an auxiliary homogeneous Dirichlet condition there. The positivity of $\widehat P_k$ at the horizon and the bounds on $\mathcal V_{k,\mathrm{rem}}$ stated after Eq.~\eqref{eq:45}, which are uniform in $k$, allow the same inner radius for every $k\ne0$.

Near $x=0$, the transformation $H=\sqrt{\frac{\ell_\alpha^2w}{d_0x}}\,U$ gives Eq.~\eqref{eq:53} with $n=1$, corresponding to the radial Laplacian on a disk plus a smooth potential. On the auxiliary disk times the angular circle, $c_\theta^2>0$ makes the principal spatial operator elliptic, and the remainder in Eq.~\eqref{eq:45} is bounded uniformly in $k$, together with each fixed radial derivative. Regularity at the disk centre and the inner Dirichlet condition define a self-adjoint comparison operator. Smooth compact initial data belong to the domains of all its powers. The wave group preserves these domains; elliptic estimates and commuting angular derivatives give a smooth solution with convergent Fourier sums. Equation~\eqref{eq:71} then supplies smooth metric and scalar perturbations satisfying $\delta E_{ab}=0$ and $\delta E_\phi=0$, the linearizations of Eqs.~\eqref{eq:4}--\eqref{eq:5}.

To establish finite propagation speed in the radial direction, use angular $L^2$ norms and let $\mathsf V_{\rm rem}$ act on harmonic $k$ by multiplication by $\mathcal V_{k,\mathrm{rem}}$. Then
\begin{equation}
\begin{gathered}
\mathcal E=\frac w2\left(\|U_\tau\|^2+\|U_x\|^2
+c_\theta^2\|U_\theta\|^2+\mu_0^2\|U\|^2\right),
\qquad j=w\operatorname{Re}\langle U_\tau,U_x\rangle,\\
\partial_\tau\mathcal E-\partial_xj
=w\operatorname{Re}\langle U_\tau,(\mu_0^2-\mathsf V_{\rm rem})U\rangle
\le C\mathcal E,\qquad |j|\le\mathcal E.
\end{gathered}
\label{eq:58}
\end{equation}
Integration over $0<x<d-\tau$ gives nonpositive flux at the moving boundary. Since the initial energy there vanishes, $U=0$ for $x+\tau<d$. Equation~\eqref{eq:71} gives zero metric and scalar perturbations in the same region. The original sources are therefore preserved for $\tau<d$. This inner Dirichlet condition is only an auxiliary choice for the comparison solution.

For the upper bound, suppose that a solution preserving the original sources exists to $T>d$. Compact initial support and zero initial velocity give $b_k(0)=\partial_\tau b_k(0)=0$ for every nonzero angular harmonic. Equation~\eqref{eq:51} therefore gives
\begin{equation}
\left(\partial_\tau^2+\frac{k^2}{\ell_\alpha^2}\right)b_k=0,
\qquad b_k(0)=\partial_\tau b_k(0)=0.
\end{equation}
Uniqueness for this ordinary differential equation implies $b_k(\tau)=0$ for $0\le\tau<T$.

Equation~\eqref{eq:41} has time-independent coefficients and contains only second time derivatives. Since the initial velocity vanishes, its solution extends evenly through $\tau=0$. Choose $d<T_0<\min(T,x_h)$. For each harmonic, the transformed field $H=\sqrt{\frac{\ell_\alpha^2w}{d_0x}}\,U_k$ has vanishing boundary value on $|\tau|<T_0$, and Eqs.~\eqref{eq:49} and \eqref{eq:79} give $H_x(\tau,0)=0$. Applying the boundary uniqueness estimate in Eq.~\eqref{eq:55} yields
\begin{equation}
U_k(\tau,x)=0
\qquad\text{for }0<x,\quad x+|\tau|<T_0.
\end{equation}

In particular, every initial Fourier coefficient vanishes for $0<x<T_0$. Since the initial displacement has zero angular mean, the full initial field vanishes on this interval. This contradicts the definition of $d$ as the distance from the boundary to its support, because $T_0>d$. The contradiction also applies to smooth packets that vanish to every order at the edge of their support. The triangle used in the estimate lies entirely outside the metric horizon, so the argument does not depend on data entering from the interior. Together with the lower bound, this proves Eq.~\eqref{eq:57}.

For the circular sector, fixing the mass gives $u_0=0$, and fixing the finite scalar source gives $u_2=0$ in Eq.~\eqref{eq:52}. The transformed field $H=\sqrt{\frac{d_0x}{\ell_\alpha^2Dr}}\,\frac{U_c}{x^2}$ therefore has boundary value $H(\tau,0)=u_2(\tau)=0$. Its regularity and the zero initial velocity permit the same even extension and boundary uniqueness argument, now with $n=3$.

\subsection{Energy balance with prescribed incoming data}

We use the following positive quadratic form to estimate solutions of Eq.~\eqref{eq:41} on $0\le x\le x_h$. The total canonical energy also depends on boundary contributions. The coefficients in Eq.~\eqref{eq:44} satisfy
\begin{equation}
\widehat P_k\ge2X_hk^2,
\qquad 0\le\widehat\nu\le\frac{4\alpha DXr}{B},
\qquad \widehat\nu=O(x).
\label{eq:80}
\end{equation}
Use $\cos(k\theta)/\sqrt\pi$ for $k>0$ and $\sin(|k|\theta)/\sqrt\pi$ for $k<0$, so the sum includes both independent real harmonics. Define
\begin{equation}
E_{\rm def}=\frac{\alpha}{4\pi}
\sum_{k\ne0}\int_0^{x_h}w\left[|U_{k,\tau}|^2+
|U_{k,x}-\widehat\nu U_k|^2+\widehat P_k|U_k|^2\right]dx.
\label{eq:81}
\end{equation}

The bulk Hamiltonian obtained directly from Eq.~\eqref{eq:42} satisfies
\begin{align}
E_{\rm bulk}
&=\frac{\alpha}{4\pi}\sum_{k\ne0}\int_0^{x_h}w
\left(|U_{k,\tau}|^2+|U_{k,x}|^2+\mathcal V_k|U_k|^2\right)dx,\\
E_{\rm bulk}
&=E_{\rm def}+\frac{\alpha}{4\pi}\sum_{k\ne0}
\left[w\widehat\nu|U_k|^2\right]_0^{x_h},\\
\frac{dE_{\rm bulk}}{d\tau}
&=\frac{\alpha}{2\pi}\sum_{k\ne0}
\left[w\operatorname{Re}(\overline{U_{k,\tau}}U_{k,x})\right]_0^{x_h}.
\end{align}
The AdS terms vanish because $w=O(x)$, $\widehat\nu=O(x)$ and $U_{k,x}=O(x)$. The term $\frac{\alpha}{4\pi}\sum_{k\ne0}[w\widehat\nu|U_k|^2]_{x=x_h}$ remains in $E_{\rm bulk}-E_{\rm def}$. Subtracting its time derivative from $dE_{\rm bulk}/d\tau$ replaces $U_{k,x}$ by $U_{k,x}-\widehat\nu U_k$ in the flux, giving Eq.~\eqref{eq:83}. These energies are obtained from the bulk action for $U_k$; a canonical energy for a specified boundary variational problem must also include its boundary contributions. Conservation requires zero net flux under that prescription. The estimate assumes that a solution exists.

Equations~\eqref{eq:44} and \eqref{eq:80} also give $\widehat P_k\le Ck^2$ with $C$ independent of $k$. Thus $E_{\rm def}$ is uniformly equivalent to the square of the weighted norm of the time, radial and angular derivatives. With incoming and outgoing combinations
\begin{equation}
g_k=(U_{k,\tau}+U_{k,x}-\widehat\nu U_k)_h,
\qquad o_k=(U_{k,\tau}-U_{k,x}+\widehat\nu U_k)_h,
\label{eq:82}
\end{equation}
differentiating Eq.~\eqref{eq:81} and using Eqs.~\eqref{eq:41} and \eqref{eq:44} cancels the volume terms. Equation~\eqref{eq:79} sets the AdS flux to zero, leaving
\begin{equation}
E_{\rm def}(T)+\frac{\alpha w_h}{8\pi}
\sum_k\int_0^T|o_k|^2d\tau
=E_{\rm def}(0)+\frac{\alpha w_h}{8\pi}
\sum_k\int_0^T|g_k|^2d\tau.
\label{eq:83}
\end{equation}
The term $-\widehat\nu U_k$ is part of the prescribed incoming datum $g_k$ in Eq.~\eqref{eq:82}. Applying Eq.~\eqref{eq:83} to the difference of two solutions gives uniqueness and continuous dependence whenever solutions with the prescribed metric and scalar boundary values exist.

For two solutions with the same exterior initial data, Eq.~\eqref{eq:58} on $0<x<x_h-\tau$ gives equality of the two solutions in
\begin{equation}
\tau\ge0,\qquad 0<x<x_h,\qquad \tau+x<x_h.
\label{eq:84}
\end{equation}
A disturbance from the interior can reach radius $r$ only after optical time $x_h-x(r)$; smooth outward pulses can reach this bound before contact with the AdS boundary.

For circular perturbations at fixed mass, Eq.~\eqref{eq:37} gives
\begin{equation}
\begin{aligned}
E_c(T)
+\frac{W_{c,h}}{32q}\int_0^T
\left|(U_{c,\tau}-U_{c,x})_h\right|^2d\tau
={}&E_c(0)\\
&+\frac{W_{c,h}}{32q}\int_0^T
\left|(U_{c,\tau}+U_{c,x})_h\right|^2d\tau.
\end{aligned}
\label{eq:88}
\end{equation}
For the regular branch, the energy flux at the AdS boundary vanishes. The fixed scalar boundary values still require $u_2=0$ in Eq.~\eqref{eq:52}.

\bibliographystyle{unsrt}
\bibliography{ref}
\end{document}